\documentclass[aps,twocolumn,amsmath,amssymb,preprintnumbers]{revtex4}
\usepackage{amsmath} \usepackage{amsfonts} \usepackage{amssymb}
\usepackage{bbm}
\usepackage{epsfig}
\usepackage{graphics}
\usepackage{graphicx}
\newcommand{\be}{\begin{equation}}
\newcommand{\ee}{\end{equation}}
\newcommand{\ba}{\begin{eqnarray}}
\newcommand{\ea}{\end{eqnarray}}
\newcommand{\nn}{\nonumber}
\newcommand{\kr}{\rangle}
\newcommand{\kl}{\langle}
\newcommand{\h}{{\cal H}}
\newcommand{\D}{{\cal D}}
\newcommand{\cl}{{\cal L}}
\newcommand{\z}{\tilde z_}
\newcommand{\zz}{\tilde z^}
\newcommand{\x}{\tilde x_}

\newcommand{\y}{(\tilde y)}
\newcommand{\mn}{_{\mu\nu}}
\newcommand{\hmn}{^{\mu\nu}}
\newcommand{\hl}{\hat{\cal L}}
\newcommand{\hp}{\hat\partial_}

\begin{document}

\title[ ]{Lattice diffeomorphism invariance}

\author{C. Wetterich}
\affiliation{Institut  f\"ur Theoretische Physik\\
Universit\"at Heidelberg\\
Philosophenweg 16, D-69120 Heidelberg}

\begin{abstract}
 
We propose a lattice counterpart of diffeomorphism symmetry in the continuum. A functional integral for quantum gravity is regularized on a discrete set of space-time points, with fermionic or bosonic lattice fields. When the space-time points are positioned as discrete points of a continuous manifold, the lattice action can be reformulated in terms of average fields within local cells and lattice derivatives. Lattice diffeomorphism invariance is realized if the action is independent of the positioning of the space-time points. Regular as well as rather irregular lattices are then described by the same action. Lattice diffeomorphism invariance implies that the continuum limit and the quantum effective action are invariant under general coordinate transformations - the basic ingredient for general relativity. In our approach the lattice diffeomorphism invariant actions are formulated without introducing a metric or other geometrical objects as fundamental degrees of freedom. The metric rather arises as the expectation value of a suitable collective field. As examples, we present  lattice diffeomorphism invariant actions for a bosonic non-linear sigma-model and  lattice spinor gravity. 
\end{abstract}

\maketitle

\section{Introduction}
\label{Introduction}

A quantum field theory for gravity may be based on a functional integral. This is well defined, however, only if a suitable regularization can be given. At this point major obstacles arise. The central symmetry of general relativity is diffeomorphism symmetry or invariance of the action under general coordinate transformations. Using the metric or the vierbein as basic fields it has not been possible so far to define a regularized functional measure that is diffeomorphism invariant. This applies, in particular, to lattice regularizations where the space-time points are discrete for the regularized theory, apparently merging to a continuous manifold only in the ``continuum limit'' for distances very large compared to the lattice distance. A second issue concerns the unboundedness of the euclidean action in many formulations. Mathematical consistency of the functional integral, both in a euclidean or Minkowski setting, is then hard to realize. 

Different attempts for a lattice formulation of quantum gravity have tried to circumvent this problem by using degrees of freedom different from the metric or the vierbein. In Regge-Wheeler lattice gravity (cf. ref. \cite{Ha} for a recent report) the lengths of edges of simplices are used as basic degrees of freedom. A metric can be reconstructed from these geometrical objects. Other formulations of lattice gravity use different geometrical objects \cite{Am}, \cite{Rov}. Lattice spinor gravity \cite{A} proposes Grassmann variables as basic degrees of freedom. (See ref. \cite{Dia} for a formulation with additional link variables for the spin connection.) In such approaches the metric emerges as the expectation value of some suitable collective field. 

For all these approaches the challenge consists in showing that a continuum limit exists for which the effective action for the metric is one of the Einstein-Hilbert type. This property is strongly suggested if the continuum limit is diffeomorphism invariant, contains a field with transformation properties of the metric and is ``local'' in the sense that the first terms of an expansion in derivatives of the metric give a reasonable description for the effective gravity theory at long distances. Besides a possible cosmological constant the curvature scalar is the leading term in such a derivative expansion for the metric. 

If regularized lattice quantum gravity with a suitable continuum limit exists, this has an important consequence: the quantum field theory for gravity must be non-perturbatively renormalizable. In the language of functional renormalization for some form of a scale dependent effective action there are only two ways how an arbitrarily large separation of ``macroscopic scales'' (in our setting typical scales of particle physics as the Fermi scale, or even astronomical scales) from the microscopic scale characterized by the lattice distance can be achieved. Either there exists an ultraviolet fixed point with a few relevant directions corresponding to free parameters (``renormalized couplings'') of the effective macroscopic theory. Or, alternatively, the macroscopic couplings reach infinite or unphysical values when their running is extrapolated to short distances. This case typically indicates that degrees of freedom different from the macroscopic ones must be used for the renormalizable microscopic theory. Recent progress in functional renormalization for gravity \cite{Reu,Cod}, based on the the exact flow equation for the effective average action or flowing action \cite{CWRG}, suggests that the asymptotic safety scenario \cite{Wei} with an ultraviolet fixed point may be viable. 

The realization of some form of lattice diffeomorphism invariance which entails diffeomorphism symmetry in the continuum limit seems to us to be a major ingredient for a successful formulation of lattice quantum gravity. A diffeomorphism invariant quantum effective action for the metric has already the structure of general relativity. In Regge-Wheeler lattice gravity a version of lattice diffeomorphism invariance is based on the invariance of the action under a change of the lengths of edges of simplices \cite{RW}. It is not clear, however, if the problem of boundedness of the action can be overcome in this approach, and if a suitable continuum limit exists. For this reason we follow here the alternative approach where geometrical objects are not the ``fundamental'' degrees of freedom in the functional integral. Geometrical objects are then not available for a formulation of diffeomorphism invariance and an alternative has to be followed. 

In this paper we propose a general form of lattice quantum gravity where the functional integral is based on lattice fields, i.e. variables ${\cal H}(\tilde z)$ for every point $\tilde z$ of a space-time lattice. The basic fields may be bosons or fermions and they behave as ``scalars'' with respect to diffeomorphism symmetry. This makes the formulation of the measure for the functional integral straightforward. Lattice spinor gravity \cite{A} belongs to this class of models. What is needed is a lattice equivalent of diffeomorphism symmetry of the action. This ``lattice diffeomorphism invariance'' should be a special property of the lattice action that guarantees diffeomorphism symmetry for the continuum limit and the quantum effective action. Its formulation is the key topic of this paper. 

The type of models that we consider are not based on a fluctuating geometry. The fluctuating degrees of freedom are fields that ``live'' in some manifold. Since there is a priori no geometry and metric the manifold is a pure coordinate manifold and we choose cartesian coordinates $x^\mu$. This coordinate manifold should not be confounded with physical space-time - the latter emerges only once a dynamical metric is found which determines geometry and topology. In general, physical space-time will be curved, and physical distances do not coincide with the coordinate distances in ${\mathbbm R}^d$. As an example, the geometry of a sphere can be described by an appropriate metric for cartesian coordinates, where the behavior for $|x|\to\infty$ is related to the pole opposite to the one at $x=0$. (A full description can be achieved by a second coordinate patch covering the missing pole. Knowing the metric everywhere except for the missing pole the completion of topology is unique - we require that physics is not affected by removing or adding a point of a manifold. Our setting resembles is this aspect the emergence of geometry from correlation functions discussed in ref. \cite{CWgeo}.) Having at our disposal only fields on a coordinate manifold our setting is very close to standard lattice field theories as lattice gauge theories or scalar field theories where the metric does not fluctuate. 

The issue of diffeomorphism symmetry in the absence of a metric may be demonstrated for the continuum action of an abelian gauge theory. We may compare two (euclidean) actions involving the gauge field strength $F_{\mu\nu}$, 
\be\label{I1}
S_1=\int_xF\mn F_{\rho\sigma}\delta^{\mu\rho}\delta^{\nu\sigma}
\ee
and
\be\label{I2}
S_2=\int_x\varphi^2 F\mn F_{\rho\sigma}\epsilon^{\mu\nu\rho\sigma}.
\ee
The first one involves a fixed inverse metric $\delta\hmn$ and is not diffeomorphism symmetric, while the second does not involve any metric and is diffeomorphism symmetric. (A squared scalar field $\varphi^2$ is employed for preventing $S_2$ to be a total derivative.) The task of the present paper is to find a lattice equivalent of diffeomorphism symmetry for actions of the type $S_2$ (In $d$ dimensions those are constructed as integrals over $d$-forms.) We will relate this to different possibilities of positioning the abstract lattice points $\tilde z$ on the coordinate manifold.

In the continuum, the invariance of the action under general coordinate transformations states that it should not matter if fields are placed at a point $x$ or some neighboring point $x+\xi$, provided that all fields are transformed simultaneously according to suitable rules. In particular, scalar fields ${\cal H}(x)$ are simply replaced by ${\cal H}(x+\xi)$. After an infinitesimal transformation the new scalar field $\h '(x)$ at a given position $x$ is related to the original scalar field $\h(x)$ by 
\be\label{1}
\h'(x)=\h(x-\xi)=\h(x)-\xi^\mu\partial_\mu\h(x).
\ee
Diffeomorphism symmetry states that the action is the same for $\h(x)$ and $\h'(x)$. Implicitly the general coordinate transformations assume that the same rule for forming derivatives is used before and after the transformation.

We want to implement a similar principle for a lattice formulation. For this purpose we associate the abstract lattice points $\tilde z=(\zz0,\zz1,\zz2,\zz3)$, with integer $\zz\mu$, with points on a manifold. As mentioned before, we consider here a piece of ${\mathbbm R}^d$ with cartesian coordinates $x^\mu=(x^0,x^1,\dots x^{d-1})$, but we do not specify any metric a priori, nor assume its existence. A map $\tilde z\to x^\mu_p(\tilde z)$ defines the positioning of lattice points in the manifold. We can now compare two different positionings, as a regular lattice $x_p^\mu(\tilde z)=\tilde z^\mu\Delta$, or some irregular one with different coordinates $x'^{\mu}_p(\tilde z)$. In general, there are many different possible positionings for the same abstract lattice points $\tilde z$. In particular, we can compare two positionings related to each other by an arbitrary infinitesimal shift $x'^\mu_p=x^\mu_p+\xi^\mu_p(x)$. We emphasize that the notion of an {\em infinitesimal} neighborhood requires a continuous manifold and cannot be formulated for the discrete abstract lattice points $\tilde z$. 

Positioning of the lattice points on a manifold is also required for the notion of a lattice derivative. One can define the meaning of two neighboring lattice points $\tilde z_1$ and $\tilde z_2$ in an abstract sense. (This involves some type of ``incidence matrix'' but no manifold.) A lattice derivative of a field will involve the difference between field values at neighboring sites, $\h(\z1)-\h(\z2)$. For the definition of a lattice derivative $\hat\partial_\mu\h$ we need, in addition, some quantity with dimension of length. This is only provided by the positioning on the manifold. For our cartesian coordinates we use a simple definition of the lattice derivative by requiring for suitable pairs $(\tilde z_1,\tilde z_2)$ the relation
\be\label{2}
\h(\z1)-\h(\z2)=\big(x^\mu_p(\z1)-x^\mu_p(\z2)\big)\hat\partial_\mu\h,
\ee
with summation over repeated indices implied.  We select locally $d$ pairs for a model in $d$ dimensions, and solve the system of $d$ independent equations \eqref{2} for the $d$ different derivatives $\hat\partial_\mu\h$. This yields the local derivatives $\hat\partial_\mu\h(\tilde y)$ in terms of suitable cartesian distances and differences of lattice variables. 

Finally, the positioning of $\tilde z$ on a manifold is a crucial ingredient for the formulation of a continuum limit, where one switches from $\h(\tilde z)$ to $\h(x)$ and derivatives thereof.

If the lattice action is originally formulated in terms of $\h(\tilde z)$ only, its expression in terms of lattice derivatives will in general depend on the positioning, since the relation between $\h(\tilde z)$ and lattice derivatives \eqref{2} depends on the positioning. We can now state the principle of ``lattice diffeomorphism invariance''. A lattice action is lattice diffeomorphism invariant if its expression in terms of lattice derivatives does not depend on the positioning of the lattice points. For infinitesimally close positionings the lattice action is independent of $\xi_p$. (For a mathematically unambiguous definition of lattice diffeomorphism invariance we need a few further specifications. This is best done by investigating below particular examples.) 

In short, lattice diffeomorphism invariance has the simple intuitive meaning that it does not matter how the abstract lattice points are placed on the manifold. A similar interpretation is possible for diffeomorphism symmetry in the continuum: it does not matter how (scalar) fields are placed on a given coordinate manifold. This makes the correspondence between lattice diffeomorphism invariance and diffeomorphism symmetry in the continuum rather natural. (Diffeomorphism symmetry in the continuum can be viewed from different aspects, for example as a symmetry transformation among field variables. Not all these aspects find a direct correspondence in lattice diffeomorphism invariance.) 

The usual discussion of lattice theories considers implicitly a given fixed positioning, for example a regular lattice. We investigate here a much wider class of positionings. Only the comparison of different positionings allows the formulation of lattice diffeomorphism invariance in our setting. 

We will see that the continuum limit of a lattice diffeomorphism invariant action exhibits diffeomorphism symmetry. Also the quantum effective action is invariant under general coordinate transformations. This extends to the effective action for the metric which appears in our setting as the expectation value of a suitable collective field. We will show that lattice diffeomorphism invariance implies diffeomorphism symmetry for the quantum effective action of the metric in the continuum limit. 
 The gravitational field equations are therefore covariant with a similar general structure as in general relativity. 

In order to show diffeomorphism symmetry of the continuum limit and the effective action we use the concept of interpolating functions.  We define a version of partial derivatives of interpolating functions that takes into account the lack of knowledge of details of the interpolation. At the positions of lattice cells these derivatives equal the lattice derivatives. For smooth interpolating fields they coincide with the standard definition of partial derivatives. 

In this view, the lattice does not reflect a basic discreteness of space and time. It rather expresses the fact that only a finite amount of information is available in practice, and that arbitrarily accurate continuous functions are an idealization since they require an infinite amount of information. In a sense, we treat continuous functions similar to numerical simulations where they have to be specified by a finite amount of information. In our formulation diffeomorphism transformations are nothing else than the possibility to move the lattice points, where the information about the function is given by its value at these points, within a manifold. Diffeomorphism invariance is realized if the action in terms of fields and their derivatives does not notice this change in positions. 

This paper is organized as follows. In sects. \ref{Lattice diffeomorphism invariance in two dimensions}-\ref{Lattice spinor gravity} we discuss lattice diffeomorphism invariance in two dimensions. The particular features of two-dimensional gravity do not play a role for the issue of diffeomorphism invariance. The discussion in these sections can be extended in a straightforward way to four dimensions. Notation and geometric visualization are simplest in two dimensions, however. In sect. \ref{Lattice diffeomorphism invariance in two dimensions} we introduce our concept of lattice diffeomorphism invariance for a scalar field theory, typically a non-linear $\sigma$-model. In the following sections \ref{Interpolating fields}-\ref{Effective action for gravity and gravitational field equations} we show that lattice diffeomorphism invariance induces diffeomorphism symmetry for the continuum limit of the ``classical action'', and, most important, for the quantum effective action. 

In sect. \ref{Interpolating fields} we introduce interpolating fields and a suitable definition of derivatives as a central tool for this argument. The interpolation fields are used in sect. \ref{Diffeomorphism symmetry of continuum action} in order to establish the diffeomorphism symmetry of the continuum limit of the classical action. Sect. \ref{Effective action} turns to the quantum effective action for the scalar fields and establishes its diffeomorphism symmetry in the continuum limit. In sect. \ref{Metric} we introduce the metric as the expectation value of a collective field. The quantum effective action for the metric, its diffeomorphism symmetry and the covariance of the gravitational field equations are discussed in sect. \ref{Effective action for gravity and gravitational field equations}.

In sect. \ref{Lattice spinor gravity} we extend our discussion to lattice spinor gravity in two dimensions. Four dimensional models, both for fundamental scalars and fermions, are discussed in sect. \ref{Lattice diffeomorphism invariance in four dimensions}. This section is kept short since besides a more involved algebra no new concepts are needed. We draw our conclusions in sect. \ref{Conclusions}. 

\section{Lattice diffeomorphism invariance in two dimensions}
\label{Lattice diffeomorphism invariance in two dimensions} 

Basic construction principles of a lattice diffeomorphism invariant action can be understood in two dimensions. We label abstract lattice points by two integers $\tilde z=(\tilde z^0,\tilde z^1)$, with $\tilde z^0+\tilde z^1$ odd. We first consider models with scalar fields. We will be mainly interested in continuous fields, but the basic issue of positioning independence can already be understood for discrete fields.

Our simplest example employs three real bosonic discrete lattice fields $\h_k(\tilde z)=\pm 1~,~k=1,2,3$. For the formulation of the functional integral we define the measure as a product of independent sums
\be\label{3}
\int \D\h=\prod_{\tilde z}\prod_k\sum_{\h_k(\tilde z)=\pm 1}.
\ee
This measure does obviously not depend on the positioning. For a finite number of lattice points the partition function is a finite sum 
\be\label{4}
Z=\int \D\h\exp(-S),
\ee
and therefore well defined for finite $S$. 

We define the action as a sum over local cells located at $\tilde y=(\tilde y^0,\tilde y^1)$, with $\tilde y^\mu$ integer and $\tilde y^0+\tilde y^1$ even,
\be\label{5}
S=\sum_{\tilde y}\cl(\tilde y).
\ee
Each cell consists of four lattice points that are nearest neighbors of $\tilde y$, denoted by $\tilde x_j(\tilde y),j=1\dots 4$. Their lattice coordinates are 
$\tilde z\big(\tilde x_1(\tilde y)\big)=(\tilde y^0-1,\tilde y^1)~,~\tilde z\big(\tilde x_2(\tilde y)\big)=(\tilde y^0,\tilde y^1-1)~,~\tilde z\big(\tilde x_3(\tilde y)\big)=(\tilde y^0,\tilde y^1+1)$, and $\tilde z\big(\tilde x_4(\tilde y)\big)=(\tilde y^0+1,\tilde y^1)$. The local term $\cl(\tilde y)$ involves lattice fields on the four sites of the cell that we denote by $\h_k(\tilde x_j)$. We choose 
\ba\label{6}
&&\cl(\tilde y)=\frac{\alpha}{48}\epsilon^{klm}\big[\h_k(\tilde x_1)+\h_k(\tilde x_2)+\h_k(\tilde x_3)+\h_k(\x4)\big]\nn\\
&&\times\big[\h_l(\x4)-\h_l(\x1)\big]
[\h_m(\x3)-\h_m(\x2)\big]+c.c.,
\ea
such that the action specifies a general Ising type model. (The complex conjugate is omitted for real fields and is needed only for the complex fields discussed later.) At this point no notion of a manifold is introduced. We specify only the connectivity of the lattice by grouping lattice points $\tilde z$ into cells $\tilde y$ such that each cell has four points and each point belongs to four cells. This defines neighboring cells as those that have two common lattice points. Neighboring lattice points belong to at least one common cell.

Our setting can be extended to continuous fields, for example by replacing the constraint $\h_k(z)=\pm 1$ by $\Sigma_k\h^2_k(z)=1$. This is a generalized non-linear sigma-model or Heisenberg model. The partition function is symmetric with respect to global $SO(3)$-rotations between the three components $\h_k$, provided we replace the functional measure \eqref{3} by the standard $SO(3)$-invariant measure for every $\tilde z$. A generalization to unbounded fields, e.g. $-\infty<\h_k(\tilde z)<\infty$, is more problematic because of the unboundedness of the action \eqref{6} even for a finite number of lattice sites. A further generalization uses complex fields $\h_k$ with constraint $\h^*_k(\tilde z)\h_k(\tilde z)=1$. For a suitable invariant measure the action and partition function are invariant under unitary transformations of the group $SU(3)$. The action \eqref{5}, \eqref{6} is real and we concentrate on real $\alpha$. 

We now proceed to an (almost) arbitrary positioning of the lattice points on a piece of ${\mathbbm R}^2$ by specifying positions $x^\mu_p(\tilde z)$. This associates to each cell a ``volume'' $V(\tilde y)$,
\be\label{7}
V(\tilde y)=\frac12\epsilon_{\mu\nu}(x^\mu_4-x^\mu_1)(x^\nu_3-x^\nu_2),
\ee
with $\epsilon_{01}=-\epsilon_{10}=1$ and $x^\mu_j$ shorthands for the positions of the sites $\tilde x_j$ of the cell $\tilde y$, i.e. $x^\mu_j=x^\mu_p\Big(\tilde z\big(\x{j}(\tilde y)\big)\Big)$. The volume corresponds to the surface inclosed by straight lines joining the positions of the four lattice points $\x j(\tilde y)$ of the cell in the order $\x1,\x2,\x4,\x3$. For simplicity we restrict the discussion to ``deformations'' of the regular lattice, $x^\mu_p=\tilde z^\mu\Delta$, where $V(\tilde y)$ remains always positive and the path of one point during the deformation never touches another point or a straight line between two other points at the boundary of the surface. We use the volume $V(\tilde y)$ for the definition of an integral over the relevant region of the manifold
\be\label{8}
\int d^2 x=\sum_{\tilde y}V(\tilde y),
\ee
where we define the region by the surface covered by the cells appearing in the sum.

We next express the action \eqref{5}, \eqref{6} in terms of average fields in the cell
\be\label{9}
\h_k(\tilde y)=\frac14\sum_j\h_k\big(\x{j}(\tilde y)\big)
\ee
and lattice derivatives \eqref{2} associated to the cell
\ba\label{9A}
\hat\partial_0\h_k\y&=&\frac{1}{2V\y}\Big\{(x^1_3-x^1_2)\big(\h_k(\x4)-\h_k(\x1)\big)\nn\\
&&-(x^1_4-x^1_1)\big(\h_k(\x3)-\h_k(\x2)\big)\Big\},\nn\\
\hat\partial_1\h_k\y&=&\frac{1}{2V\y}\Big\{(x^0_4-x^0_1)\big(\h_k(\x3)-\h_k(\x2)\big)\nn\\
&&-(x^0_3-x^0_2)\big(\h_k(\x4)-\h_k(\x1)\big)\Big\}.
\ea
For the pairs $(\x{j_1},\x{j_2})=(\x4,\x1)$ and $(\x3,\x2)$ the lattice derivatives obey 
\be\label{10}
\h_k(\x{j_1})-\h_k(\x{j_2})=
(x^\mu_{j_1}-x^\mu_{j_2})\hat\partial_\mu\h_k(\tilde y).
\ee
In terms of cell average and lattice derivatives all quantities in $\cl(\tilde y)$ depend on the cell variable $\tilde y$ or the associated position of the cell $x^\mu_p(\tilde y)$ that we take somewhere inside the surface of the cell, the precise assignment being unimportant at this stage. In this form we denote $\cl\y$ by $\hat\cl(\tilde y;x_p)$ or $\hat\cl(x;x_p)$, where $\hl(x)$ only depends on quantities with support on discrete points in the manifold corresponding to the cell positions. We indicate explicitly the dependence on the choice of the positioning by the argument $x_p$. 

The action appears now in a form referring to the positions on the manifold
\be\label{11}
S(x_p)=\int d^2x\bar\cl(\tilde y;x_p)=\int d^2x\bar\cl(x;x_p),
\ee
with 
\be\label{12}
\bar\cl(\tilde y;x_p)=\bar\cl(x;x_p)=\frac{\hat\cl(\tilde y;x_p)}{V(\tilde y;x_p)}.
\ee
Lattice diffeomorphism invariance states that for fixed $\h(\tilde y)$ and $\hat\partial_\mu\h(\tilde y)$ the ratio $\bar\cl(\tilde y;x_p)$ in eq. \eqref{12} is independent of the positioning, or independent of $\xi_p$ for infinitesimal changes of positions $x'_p=x_p+\xi_p$,
\be\label{12A}
\bar\cl(\tilde y;x_p+\xi_p)=\bar\cl(\tilde y;x_p)~,~S(x_p+\xi_p)=S(x_p).
\ee
This definition provides the necessary specification for the precise meaning of lattice diffeomorphism invariance.

The $\xi_p$-independence of $\bar\cl(\tilde y;x_p)$ means that the dependence of $V(\tilde y;x_p)$ and $\hat\cl(\tilde y;x_p)$ on the positioning $x_p$ must cancel. Inserting eqs.~\eqref{9},~\eqref{10} in eq. \eqref{6} yields
\be\label{13}
\hat\cl\y=\frac{\alpha}{12}\epsilon^{klm}V\y\h_k\y\epsilon^{\mu\nu}\hat\partial_\mu\h_l\y\hat\partial_\nu\h_m\y+c.c.,
\ee
and we find indeed that the factor $V\y$ cancels in $\bar\cl\y=\hl\y/V\y$. Thus the action \eqref{5}, \eqref{6} is  lattice diffeomorphism invariant. This property is specific for a certain class of actions - for example adding to $\epsilon^{klm}$ a quantity $s^{klm}$ which is symmetric in $l\leftrightarrow m$ would destroy lattice diffeomorphism symmetry. 

For all typical lattice theories the formulation of $\cl\y$ only in terms of next neighbors and common cells (not using a distance) does not refer to any particular positioning. However, once one proceeds to a positioning of the lattice points and introduces the concept of lattice derivatives, the independence on the positioning of $\bar\cl(\tilde y;x_p)$ for fixed $\h\y$ and $\hat\partial_\mu\h\y)$ is not shared by many known lattice theories. For example, standard lattice gauge theories are not lattice diffeomorphism invariant. This is also the basic difference of the actions $S_1$ and $S_2$ (eqs. \eqref{I1}, \eqref{I2}) mentioned in the introduction: only for $S_2$ it is possible to find a lattice diffeomorphism invariant action such that it is obtained in the continuum limit. 

We may also introduce higher lattice derivatives and express
\ba\label{15A}
&&\h_k(\x1)-\h_k(\x2)-\h_k(\x3)+\h_k(\x4)={\cal E}\h_k(\tilde y)\nn\\
&&=c_1(x)(\partial^2_0-\hat\partial^2_1)\h_k(\tilde y)+c_2(x)\hat\partial_0\hat\partial_1\tilde\h_k(\tilde y).
\ea
The functions $c_1$ and $c_2$ depend on the positions $x_j$ of the cell points $\x{j}$. (For the regular lattice one has $c_1=\Delta^2,c_2=0$.) Using eqs. \eqref{9}, \eqref{10}  we can write the lattice fields $\h_k(\tilde z)$ in terms of ${\cal K}_{s,k}(\tilde y)=\{\h_k\y),\hat\partial_0\h_k(\tilde y)$, $\hat\partial_1 \h_k\y,(\hat\partial^2_0-\hat\partial^2_1)\h_k\y,\hat\partial_0\hat\partial_1\h_k\y\},s=1\dots 5$. One can therefore express an arbitrary ${\cal L}\y$ in terms of lattice derivatives. Lattice diffeomorphism invariance is realized if $V\y{\cal L}\y$, once expressed in terms of ${\cal K}_{s,k}\y$, does not depend on the positions $x_p(\tilde z)$. We note that the lattice variables ${\cal K}_{s,k}\y$ are not independent. First, for a given positioning only one particular linear combination of ${\cal K}_4$ and ${\cal K}_5$ appears and we may set the orthogonal one to zero. Second, a given $\h(\tilde z)$ can be expressed in terms of ${\cal K}_s\y$ for all four cells $\tilde y$ to which $\tilde z$ belongs. This constraint relates variables ${\cal K}_s\y$ in different cells to each other. Imposing these constraints one may consider the change from $\h(\tilde z)$ to ${\cal K}_s\y$ as a change of basis for the lattice variables. For our particular action \eqref{6} the higher derivatives ${\cal K}_4$ and ${\cal K}_5$ are absent. In this case we can restrict $s=1\dots 3$. 

From this perspective we consider a family of basis changes for the lattice variables $\h(\tilde z)\to {\cal K}_s\y$. The specific basis ${\cal K}_s\y$ is determined by a specific positioning of the lattice points. A repositioning of the lattice points can be considered as a map between two members of this family ${\cal K}^{(1)}_s\y\to{\cal K}^{(2)}_2\y$. Lattice diffeomorphism symmetry states that the action remains invariant with respect to this map if the change of the cell volume is taken into account properly. These aspects are discussed in more detail in the appendix.

\section{Interpolating fields} 
\label{Interpolating fields}

As a key result of this note we state that the continuum limit of both the action and the quantum effective action is diffeomorphism symmetric if the lattice action is lattice diffeomorphism invariant.  We will specify the precise meaning of this statement and detail our argument in the following. A central ingredient is the observation that diffeomorphism transformations can be realized by repositionings of the lattice variables, without transforming the lattice variables themselves. 

We first want to show that the continuum limit of a lattice diffeomorphism invariant model exhibits diffeomorphism symmetry. The continuum limit will be defined in terms of interpolating functions. Consider a system of functions $f(x)$ that are completely determined by their values $f_n=f(x_n)$ at ${\cal N}$ points $x_n$ in ${\mathbbm R}_2$, while they interpolate in some specified way inbetween those points. We take ${\cal N}$ to be equal to the number of lattice points. We further define which points are neighbors in a way that allows a two-dimensional ordering similar to the lattice points $\tilde z$. Beyond the notion of neighborship the position of the points $x_n$ in ${\mathbbm R}_2$ is arbitrary. We identify the points $x_n$ with the arbitrary positions $x_p(\z n)$ of the ${\cal N}$ lattice points $\z n$ in ${\mathbbm R}_2$. The information contained in some lattice field $\tilde f(\tilde z)$ is equivalent to the values $\{f_n\}$, with $f_n=\tilde f(\z n)$. 

The map between the lattice functions $\tilde f(\tilde z)$ and the interpolating fields $f(x)$ depends both on the choice of interpolation and on the selection of points $x_n=x_p(\tilde z_n)$. This reflects the fact that a complete specification of the interpolating function $f(x)$ requires the specification of (i) the points $x_n$, (ii) the values $f(x_n)$, and (iii) an interpolation description. Even for a given prescription of interpolation the function $f(x)$ depends on the locations $x_n$ where it takes the values $f_n$. For a given lattice function $\tilde f(\tilde z)$ two different positionings of the lattice points in ${\mathbbm R}_2$ will lead to two different interpolating functions $f(x)$. 

In contrast to the values of $f$ at the points $x_n$ the information about the behavior of $f(x)$ for $x\neq x_n$ is not contained in the values of the lattice field $\tilde f(\z n)$. It depends on the specific interpolation which is chosen. This interpolation may be continuous and differentiable, or only piecewise differentiable in cells or parts of cells. For the lattice regularized theory nothing should depend on the particular choice of interpolation since the latter involves information not available in the lattice model. In this sense space becomes ``fuzzy''. All interpolations should represent the same physics.

For a given choice of interpolation we next define the notion of an ``interpolation derivative'' for an interpolating function $f(x)$. We select two vectors fields $\eta^\mu_1(x),\eta^\mu_2(x)$ that are nowhere parallel or nonvanishing, $\epsilon\mn\eta^\mu_1(x)\eta^\nu_2(x)\neq 0$. Furthermore, we introduce two additional fields $\zeta^\mu_1(x),\zeta^\mu_2(x)$. Interpolation derivatives at $x$ are defined by the two relations $(a=1,2)$
\ba\label{A1} 
f\big (x+\zeta_a(x)+\eta_a(x)\big)-f(x+\zeta_a(x)\big)=\eta^\mu_a(x)\hat\partial_\mu f(x). \nn\\
\ea
This definition becomes the usual definition of partial derivatives for infinitesimal $\eta_a$ and $\zeta_a$. (In this limit $\zeta$ drops out since $\partial_\mu f(x)$ and $\partial_\mu f(x+\zeta)$ are not different for infinitesimal $\zeta$.) In contrast to the usual differentiation we do not take here the vectors $\eta_a,\zeta_a$ to be infinitesimal but keep finite values. This is justified by the ``fuzziness'' of space, since for infinitesimal $\eta_a,\zeta_a$ the derivatives would strongly depend on the chosen interpolation. Of course, the interpolation derivatives $\hat\partial_\mu f(x)$ will now depend on the particular choice of $\eta_a(x)$ and $\zeta_a(x)$. 

Let us consider the particular case where for each position of a lattice cell $x=x_p(\tilde y)$ the vectors $\eta_a$ correspond to the diagonals in the associated cell $\tilde y$. More specifically, we consider
\ba\label{A2}
\eta_1(x)&=&x_p\big(\tilde z\big(\x4(\tilde y)\big)\big)-x_p\big(\tilde z\big(\x1(\tilde y)\big)\big),\nn\\
\eta_2(x)&=&x_p\big(\tilde z\big(\x3(\tilde y)\big)\big)-x_p\big(\tilde z\big(\x2(\tilde y)\big)\big),\nn\\
\zeta_1(x)&=&x_p\big(\tilde z\big(\x1(y)\big)\big)-x_p(\tilde y),\nn\\
\zeta_2(x)&=&x_p\big(\tilde z\big(\x2(\tilde y)\big)\big)-x_p(\tilde y).
\ea
With this choice the interpolation derivatives at $x=x_p\big(\tilde y)$ can be expressed in terms of two differences of values of the lattice field in the cell $\tilde y$, namely between positions $\x4$ and $\x1$ or $\x3$ and $\x2$, respectively. By virtue of eq. \eqref{10} the interpolation derivatives are given by the lattice derivatives
\be\label{A3}
\hat \partial_\mu f(x)=\hat\partial_\mu\tilde f(\tilde y),
\ee
where $\tilde y$ denotes the cell associated to $x$. Eq. \eqref{A2} fixes the vectors $\eta_a$ and $\zeta_a$ for all positions $y_n=x_p(\tilde y_n)$ of the ${\cal N}$ cells $\tilde y_n$. For values of $x$ inbetween the points $y_n$ we may again choose some interpolation for $\eta_a(x),\zeta_a(x)$. The values of $\hat\partial_\mu f(x)$ for $x\neq y_n$ will depend on the interpolation.

The choice \eqref{A2} is particular since only information contained in a lattice field is required for the computation of interpolation derivatives at the points $y_n$. On the other hand, there are other choices for $\eta_a(x)$ which allow the computation of $\hat\partial_\mu f(x)$ in terms of the information contained in a lattice function $\tilde f(\tilde z)$. For example, we may replace $\eta_1(x)$ by 
\ba\label{A3a}
\eta'_1(x)&=&x_p\big(\tilde z\big(\x4(\tilde y')\big)\big)-x_p\big(\tilde z\big(\x1(\tilde y)\big)\big),\nn\\
\tilde y'&=&\tilde y+(2,0).
\ea
In fact, whenever both vectors $\eta_a(x)$ join two pairs of points $x_n=x_p(\tilde z_n)$ (and are not parallel), the derivatives $\hat\partial_\mu f(x)$ can be expressed in terms of the lattice field $\tilde f(\tilde z)$ associated to to $f(x)$. (We discuss here $x=x_p(\tilde y)$ and keep the same $\zeta_a(x)$ as before.)

We will now focus on smooth lattice fields $\tilde f(\tilde z)$ characterized by the property that the lattice derivatives $\hat\partial_\mu \tilde f(\tilde y)$ are the same for all cells $\tilde y$ within a region ${\cal R}$, up to small higher order corrections. (This can be realized for the generalized non-linear $\sigma$-model, not for the generalized Ising model.) In this case the interpolation derivatives $\hat\partial_\mu f(x)$ become independent of the choice of $\eta_a$, again up to small corrections. (This holds provided that $x,x+\zeta_a$ and $x+\zeta_a+\eta_a$ all belong to the region ${\cal R}$. We concentrate here on those $x$ and $\eta_a$ for which $\hat\partial_\mu f$ can be expressed in terms of the lattice fields $\tilde f(\tilde z)$ as discussed above.) Indeed, we note for $\eta'_1$ in eq. \eqref{A3a} the relation
\be\label{A4}
\eta'_1(x)=\eta_1(x)+\eta_1\big (x+\eta_1(x)\big).
\ee
Defining the derivatives with the set of vectors $\eta'_1,\eta_2$ amounts to use for eq. \eqref{A1} $\big($ with $x+\eta_1=x+\eta_1(x),\tilde x=x+\zeta_1(x)\big)$
\ba\label{A5}
&&f\big(\tilde x+\eta_1(x)+\eta_1(x+\eta_1\big)-f(\tilde x)\nn\\
&&\quad =\big(\eta^\mu_1(x)+\eta^\mu_1(x+\eta_1)\big)\hat\partial_\mu f(x)\nn\\
&&\quad =f\big(\tilde x+\eta_1(x)+\eta_1(x+\eta_1)\big)-f(\tilde x+\eta_1)\nn\\
&&\quad \quad +f(\tilde x+\eta_1)-f(\tilde x)\nn\\
&&\quad =\eta^\mu_1(x+\eta_1)\hat\partial_\mu f(\tilde y')+\eta^\mu_1(x)\hat\partial_\mu f(\tilde y)\nn\\
&&\quad \approx \big(\eta^\mu_1(x)+\eta^\mu_1(x+\eta_1)\big)\hat\partial_\mu f(\tilde y).
\ea
We conclude that the relation \eqref{A3} also holds, up to small corrections, for the derivative formed with $(\eta'_1,\eta_2)$, which is therefore the same as the one formed with $(\eta_1,\eta_2)$. This can easily be  generalized to other vectors $\eta_a$ linking points $x_n$ within the region ${\cal R}$. The corrections $\sim\hat\partial_\mu\tilde f(\tilde y')-\hat\partial_\mu\tilde f(\tilde y)$ can be linked to higher derivatives as $\hat\partial_\mu\hat\partial_\nu f(x)$, but we will not discuss this issue here. 

Our definition of interpolation derivatives \eqref{A1} adapts the notion of a derivative to a situation where only limited information about a function is available - in our case the points $x_n$ and the values $f_n=f(x_n)$. Our notion of smooth functions adapts the concept of differentiability to this situation. On a large enough scale of coordinate distances in ${\mathbbm R}^2$ the vectors $\eta_a$ can be considered as infinitesimal, and the independence on the choice of $\eta_a$ reflects the independence on the limiting procedure for differentiable functions. On distance scales of the order $|\eta_a|$, however, this view is no longer possible and particular vectors as the ones in eq. \eqref{A2} have to be selected in order to retain the computability of the derivatives in terms of the pairs $(x_n,f_n)$. 

With this perspective the lattice regularization may not be considered as a sign of a discrete nature of space and time. It rather reflects the fact that only a limited amount of information is available for the specification of continuous functions. In short, our definitions of interpolating fields and their interpolation derivatives are chosen such that they coincide with the lattice field and lattice derivatives for all cell locations. Away from the cell locations both $f(x)$ and $\hat\partial_\mu f(x)$ are no longer computable in terms of the lattice function $\tilde f(\tilde z)$ and the positions $x_n=x_p(\z n)$. They depend on the specific choice of an interpolation. 

The definition of interpolation derivatives \eqref{A1} can also be used for discrete lattice variables, as $f_n=\pm 1$. In this case, however, the notion of smooth lattice fields is no longer available for the microscopic degrees of freedom. This will be different for the expectation values of the discrete variables which are continuous real numbers. 

In summary of this section, we have defined interpolating functions and their interpolation derivatives with the following properties: (i) The interpolating functions coincide with the lattice functions for all positions $x_n$ of lattice points. (ii) The interpolation derivatives of the interpolating functions coincide with the lattice derivatives for all cell positions $y_n$. (iii) The interpolation derivatives coincide with the usual definition of partial derivatives for smooth functions. (iv) The functional integral over interpolating functions involves integrals over the values $f_n=f(x_n)$. The precise interpolating function $f(x)$ that is specified by the set $\{f_n\}$ depends on the positions $x_n$ and on the details of the interpolation prescription. (v) For a given interpolation prescription a change of positions $\{x_n\}$ results for given $\{f_n\}$ in a change of the function $f(x)$. Such changes will be associated in sect. \ref{Diffeomorphism symmetry of continuum action} with fuzzy diffeomorphism transformations.

\section{Diffeomorphism symmetry of continuum action} 
\label{Diffeomorphism symmetry of continuum action}

Coming back to the lattice model \eqref{5}, \eqref{6} we associate interpolating fields $\h_k(x)$ to the lattice fields $\h_k(\tilde z)$. With the choice of interpolation derivatives \eqref{A1}, \eqref{A2} and eq. \eqref{13} we can define the continuum action as a functional of the interpolating fields
\be\label{x1}
S=\frac{\alpha}{12}\int d^2x\epsilon^{klm}\epsilon^{\mu\nu}
\h_k(x)\hat\partial_\mu\h_l(x)\hat\partial_\nu\h_m(x)+c.c.
\ee
For smooth fields we replace $\hat\partial_\mu\to\partial_\mu$. In this limit we associate the continuum action \eqref{x1} to the continuum limit of the lattice action \eqref{5}, \eqref{6}. 

We consider the lattice action as the basic object since it is computable in terms of the available information encoded in $\h_k(\tilde z)$. The continuum action should mainly be considered as a very useful approximation for smooth enough fields. In a certain sense the usual role of space as a manifold is realized only on distance scales large compared to the distance between lattice points. Interpolating fields and the value of the continuum action \eqref{x1} for these fields are formally defined for arbitrarily small coordinate distances $|x_1-x_2|$ in ${\mathbbm R}_2$. However, the dependence on the choice of interpolation introduces physical ambiguity on very small distance scales if no information on a particular ``physical interpolation prescription'' is available. 

We observe that the continuum limit of the action \eqref{x1} is diffeomorphism symmetric in the usual sense if the fields $\h_k(x)$ transform as scalars under general coordinate transformations \eqref{1}. Then $\partial_\mu\h_k(x)$ transforms as a vector, and the contraction with the $\epsilon$-tensor implies an invariant action. The usual notion of diffeomorphism transformations can be realized for smooth enough ``macroscopic fields'' where the derivatives $\partial_\mu\h(x)$ coincide with the standard definition of partial derivatives, while $\eta_a(x)$ as well as the displacements $\xi^\mu(x)$ in eq. \eqref{1} can be considered effectively as arbitrary infinitesimal functions of $x$. 

In a more microscopic approach the notion of independent infinitesimal displacements $\xi^\mu(x)$ for every point $x$ is no longer meaningful in view of the limited amount of information available for the interpolating functions and the associated fuzziness of space. Since the lattice information concerns only a finite number of points $x_n$, we only should consider the ${\cal N}$ infinitesimal displacements $\xi(x_n)=\xi_n$ as independent parameters of general coordinate transformations. The associated interpolating functions $\xi(x)$ can be constructed with the same interpolation prescription as for $\h(x)$. The infinitesimal displacements $\xi_n$ can then be identified with a change of positioning of the lattice points $x_p(\tilde z)\to x_p(\tilde z)+\xi_p(\tilde z)$, with $\xi_n=\xi_p(\tilde z_n)$. The displacements specified in this way by the points $x_n$ and $\xi_n=\xi(x_n)$ are associated to fuzzy diffeomorphism transformations. We will establish in this section the invariance of the continuum action under fuzzy diffeomorphisms. Before doing so, we will specify further our choice of interpolation. 

For smooth enough fields the continuum action \eqref{x1} coincides with the lattice action \eqref{5}, \eqref{6} or \eqref{11}, \eqref{13}. The derivative $\partial_\mu\h_k(x)$ can be identified with $\hat\partial_\mu\h_k(\tilde y)$ for all $x$ in a neighborhood of $x_p\y$ and $\h_k(x)$ is well approximated by the cell average $\h_k(\tilde y)$. Only for fields varying on a length scale comparable to a typical coordinate distance between lattice points the relation between the continuum action and the lattice action will depend on details of the interpolation prescription. For any specific interpolation one may express the lattice action as a functional of the interpolating fields. This ``interpolating action'', which may be rather complicated, typically differs from the continuum action \eqref{x1} for strongly varying fields. Its precise form will depend on the interpolation. It will not be relevant for smooth fields for which the difference to the continuum action disappears. 

We will discuss next a specific class of interpolations for which the interpolating action and the continuum action coincide. For this choice the functional \eqref{x1} for the interpolating fields equals precisely the lattice action \eqref{5}, \eqref{6}. For our purpose we impose the condition 
\ba\label{IA}
&&\int\limits_{\rm{cell}}d^2x\epsilon^{klm}\epsilon\hmn\h_k(x)\hat\partial_\mu\h_l(x)\hat\partial_\nu\h_m(x)\nn\\
&&=\epsilon^{klm}\epsilon\hmn V\y\h_k\y\hat\partial_\mu\h_l\y\hat\partial_\nu\h_m\y.
\ea
Interpolations obeying the condition \eqref{IA} are called ``congruent action interpolations''. It is straightforward to see that for congruent action interpolations the continuum action \eqref{x1} equals precisely the lattice action \eqref{5}, \eqref{6} for arbitrary configurations of the lattice fields $\h_k(\tilde z)$. In this case the continuum action is computable in terms of the information contained in the lattice fields without the need to know additional details of the interpolation prescription. We notice that despite of the appearance the continuum action is not invariant under continuous rotations. The interpolation derivatives in eq. \eqref{x1} are the ones that we have defined for interpolating fields by specific vectors $\eta^\mu_a,\zeta^\mu_a$. Continuous rotation symmetry arises effectively only for smooth fields. Even though the use of congruent action interpolations is not crucial for the properties of smooth fields it can be used as a convenient tool for later proofs of diffeomorphism invariance of the quantum effective action. We will use this class of interpolation functions in the following.

It may be instructive to discuss briefly a particular example for a congruent action interpolation. We choose the value of $f$ at the cell position $y=x_p(\tilde y)$ to be equal to the cell average $\tilde f(\tilde y)=\Sigma_j\tilde f(\x j)/4$, i.e. $f(y)=\tilde f\y$. Drawing straight lines between the cell position $y$ and the positions $x_j$ of the four lattice points belonging to the cell we cut the cell into four triangles. For each triangle we have now three values of $f$ at the corners, e.g. $f(x_1),f(x_2)$ and $f(y)$ for the triangle $(x_1,x_2,y)$. For all values of $x$ within this triangle we interpolate $f(x)$ by a plane in the three dimensional space spanned by the coordinates $(x^0,x^1,f)$. The orientation and position of this plane are fixed by the three values of $f(x)$ at the corners of the triangle. We apply this procedure to all four triangles of the cell. For this interpolation the values of $f(x)$ on each edge of a triangle are given by a linear interpolation between the values of $f(x)$ at the two corners bounding a given edge. Since on each boundary of the triangle $f(x)$ depends only on those two values, the interpolating function $f(x)$ is continuous on the edges of each triangle, taking the same value for neighboring triangles. This includes the edge joining two neighboring cells. Thus $f(x)$ is continuous, while $\partial_\mu f(x)$ is typically discontinuous for this type of interpolation.

Our procedure fixes the interpolation prescription up to the choice of the cell position $y=x_p(\tilde y)$. Fixing the cell position amounts to two conditions for the values of $y^0$ and $y^1$. As a first condition we impose the relation \eqref{IA}. Since the integral over interpolating functions depends on the cell position $x_p\y$ the relation \eqref{IA} can indeed be seen as a condition for $x_p\y$. Besides the condition \eqref{IA} we may impose one further condition for determining $x_p\y$. For example,we may require that the sum of the surfaces of two triangles within the cell that do not share a common side equals half the surface of the cell or comes as close as possible to this value.  

There are many other possible prescriptions for congruent action interpolations. In the following we will assume that the interpolation is continuous and differentiable everywhere in ${\mathbb R}^2$. This has the advantage that standard partial derivatives $\partial_\mu\h(x)$ are defined for the interpolating functions. (This contrasts with our example of a piecewise continuous and differentiable interpolation for which $\partial_\mu\h$ is discontinuous at the boundaries of triangles, while $\hat\partial_\mu\h$ is always well defined.) The choice of differentiable lattice congruent interpolations is not important for the continuum limit, but it will facilitate some of the formal proofs in the following. 

A condition of the type \eqref{IA} can be imposed for interpolating functions for rather arbitrary lattice actions, not necessarily being lattice diffeomorphism invariant. However, the diffeomorphism symmetry of the continuum action is a direct consequence of lattice diffeomorphism invariance. More precisely, we will see that if the lattice action is not lattice diffeomorphism invariant the continuum  action is not diffeomorphism symmetric. Thus lattice diffeomorphism invariance is a necessary condition for {\em exact} diffeomorphism symmetry of the continuum action. For the continuum limit of smooth fields this condition may be weakened. It is sufficient that possible violations of exact diffeomorphism symmetry of the continuum action vanish in the continuum limit. 

On the other hand, we will show that lattice diffeomorphism invariance implies the invariance of the continuum action under ``fuzzy diffeomorphisms''. Fuzzy diffeomorphisms are characterized by parameters $\xi^\mu_f(x)$ that are a subset of the arbitrary displacements $\xi^\mu(x)$ that characterize the general coordinate transformations. Every smooth displacement $\xi^\mu(x)$ can be well approximated by a suitable fuzzy diffeomorphism transformation $\xi^\mu_f(x)$. For the continuum limit of smooth fields and smooth displacements lattice diffeomorphism invariance therefore implies diffeomorphism symmetry of the continuum action. 

We recall in this context that ``smooth'' relates to weak variations on the scale of the lattice distance. The latter may be associated to the Planck length or even a smaller length scale. Seen from a description at a larger physical arbitrarily varying fields and displacements are smooth. We proof the statements of the preceding paragraph in the following. 

Fuzzy diffeomorphisms are defined as interpolating functions for infinitesimal repositionings of lattice points $\xi^\mu_p(\tilde z)$. Similar to the interpolating fields the fuzzy diffeomorphisms are specified at a discrete set of lattice points. At these points they equal the infinitesimal repositioning of lattice points 
\be\label{23S1}
\xi^\mu_f(x_n)=\xi^\mu_p(x_n)=x^\mu_{p,2}(\tilde z_n)-x^\mu_{p,1}(\tilde z_n),
\ee
where $x_{p,2}(\tilde z_n)$ and $x_{p,1}(\tilde z_n)$ are two infinitesimally close positions of a particular lattice point $\tilde z_n$. Inbetween the points $x_n$ the fuzzy displacements are interpolated according to some suitable interpolation description that we may choose to be the same one as for the interpolating fields. Different interpolation prescriptions lead to different $\xi^\mu_f(x)$ for $x\neq x_n$. This explains the name of ``fuzzy diffeomorphisms''. 

A given positioning of the lattice points, together with a given interpolation prescription, results in a map from the lattice field $\h(\tilde z)$ to a continuous interpolating function $\h(x)$. A general lattice action (not necessarily lattice diffeomorphism invariant) can be expressed in terms of $\h(x)$ and the interpolation derivatives $\hat\partial_\mu\h(x)$, 
\be\label{23S2}
S[\h(\tilde z)]=S[\h(x);g(x)].
\ee
In general, the coefficients in front of the different terms will depend on the positions $x_p(\tilde z)$. These ``position dependent couplings'' are denoted in eq. \eqref{23S2} by $g(x)$. Let us now consider two infinitesimally close positionings leading to two sets of fields and couplings $\big(\h_1(x),g_1(x)\big)$ and $\big(\h_2(x),g_2(x)\big)$. From eq. \eqref{23S2} one infers
\be\label{23S3}
S[\h_1(x);g_1(x)]=S[\h_2(x);g_2(x)].
\ee

The two fields $\h_2(x)$ and $\h_1(x)$ are related by a fuzzy diffeomorphism transformation
\be\label{23S4}
\h_2(x)-\h_1(x)=\delta_f\h(x)=\h(x-\xi_f)-\h(x).
\ee
In other words, a fuzzy coordinate transformation maps an interpolating field $\h_1(x)$ to another interpolating field $\h_2(x)$ which corresponds to the {\em same} lattice field $\h(\tilde z)$. The two interpolating fields $\h_1(x)$ and $\h_2(x)$ have the same values $\h_n$ at the points $x_n$, but the position of the points $x_n$ in the manifold differs. (They also obey the same interpolation prescription.) Fuzzy diffeomorphisms are maps in the space of all interpolating functions for arbitrary sets of points $\{x_n\}$. For differentiable interpolating functions $\h(x)$ one has the usual relation 
\be\label{23S5}
\delta_f\h(x)=-\xi^\mu_f(x)\partial_\mu\h(x),
\ee
where $\partial_\mu$ denotes here the usual partial derivative (not the interpolation derivative). Arbitrary smooth displacements $\xi^\mu(x)$ are approximated by a suitable $\xi^\mu_f(x)$ and we recover eq. \eqref{1}. 

A repositioning of the lattice points also changes the coupling functions $g(x)$, 
\be\label{23S6}
g_2(x)-g_1(x)=\tilde\delta_fg(x)=g(x-\xi_f)-g(x).
\ee
According to eq. \eqref{23S3} the continuum action is invariant under the simultaneous variation of fields and couplings
\be\label{23S7}
\delta_f S[\h(x);g(x)]+\tilde \delta_f S
[\h(x);g(x)]=0,
\ee
with 
\ba\label{23S8}
\delta_f S[\h(x);g(x)]=
S[\h(x-\xi_f);g(x)]-S[\h(x);g(x)],\nn\\
\tilde \delta_fS[\h(x);g(x)]=S[\h(x);g(x-\xi_f)]
-S[\h(x);g(x)].
\ea
Fuzzy diffeomorphisms act only on the fields, not on the couplings, similar to general coordinate transformations. We conclude that the continuum action is invariant under fuzzy diffeomorphisms precisely if the couplings do not depend on the positioning, $\tilde \delta_fS=0$. 

The possibility to express the continuum action in terms of $\h(\tilde x)$ without coordinate dependent couplings reflects the property of lattice diffeomorphism invariance. If $\bar\cl\y$ would show an explicit dependence on coordinates, this would show up in the continuum action (with interpolation \eqref{IA}). In analogy to the lattice action the coordinate dependence in the definition \eqref{A1} of interpolation derivatives cancels against the coordinate dependence in the volume factor. This concludes the proof of our statements concerning the relation between lattice diffeomorphism invariance and diffeomorphism symmetry of the continuum action.

\section{Effective action}
\label{Effective action}

In this section we discuss the quantum effective action for scalar fields that correspond to the expectation values of the interpolating fields $\kl \h_k(x)\kr$. We establish the (fuzzy) diffeomorphism invariance of the effective action. This follows a simple basic idea: If $\bar\cl\y$ or the associated continuum action does not notice where the lattice points are placed on the manifold, this property will also hold for the effective action. In other words, if no information on the positioning is contained in $\bar\cl \y$, the effective action will also not involve such information. Or, in an equivalent view, if the continuum action $S[\h(x)]$ does not involve position dependent couplings there is no way how a position dependence of couplings could arise on the level of the effective action. This simple result is crucial since it guarantees diffeomorphism symmetry of the quantum effective action in the continuum limit. We will show in sect. \ref{Effective action for gravity and gravitational field equations} that it generalizes to the effective action for the metric, thereby providing the basic ingredient for general relativity. 

The generating functional for the connected Greens functions of lattice variables is defined in the usual way by introducing sources
\be\label{22O}
W\big[J(\tilde z)\big]=\ln\int\D\h\exp\big\{-S+\sum_{\tilde z}\big(\h_k(\tilde z) J^*_k(\tilde z)+c.c\big)\big\},
\ee
with 
\be\label{22P}
\frac{\partial W}{\partial J^*_k(\tilde z)}=\kl\h_k(\tilde z)\kr=h_k(\tilde z).
\ee
(We omit the complex conjugate source term for real $\h_k$.) In the continuum limit the source term becomes 
\be\label{22Q}
\sum_{\tilde z}\h_k(\tilde z) J^*_k(\tilde z)=\int_x\h_k(x)j^*_k(x).
\ee
One also may define 
\be\label{22R}
\Gamma[h,J]=-W[J]+\sum_{\tilde z}\big(h_k(\tilde z) J^*_k(\tilde z)+c.c.\big),
\ee
which becomes the usual quantum effective action $\Gamma[h]$ (generating functional of $1$PI-Greens functions) if we solve eq. \eqref{22P} for $J^*(\tilde z)$ as a functional of $h(\tilde z)$ and insert this solution into eq. \eqref{22R},
\be\label{43A1}
\Gamma[h]=\Gamma\big[h,J[h]\big].
\ee
The (``functional'') derivative of the effective action with respect to $h$ yields the exact quantum field equation 
\be\label{43A2}
\frac{\partial\Gamma}{\partial h_k(z)}=J^*_k(z).
\ee

We next use an interpolating field $j(x)$ that coincides with $j_n=j(x_n)=j(\tilde z)$ for all positions of lattice points $x_n=x_p(\tilde z_n)$. Here the relation between $j(\tilde z)$ and $J(\tilde z)$ involves an approximate volume factor according to eq. \eqref{8}. The interpolation prescription for $j(x)$ is chosen such that eq. \eqref{22Q} holds for arbitrary interpolating fields and sources (not necessarily smooth), similar to eq. \eqref{IA}. The expectation value $h(x)$ is given by the same congruent action interpolation for the lattice field $h(\tilde z)$ as used for relating $\h(x)$ to $\h(\tilde z)$, such that 
\be\label{40AX}
\sum_{\tilde z}h_z(\tilde z)J^*_k(\tilde z)=\int_x h_k(x)j^*_k(x).
\ee
We are interested in $\Gamma[h(x)]$ as a functional of the interpolating fields $h(x)$. The lattice field equation \eqref{43A2} is then transmuted to an equation for continuous fields which involves a functional derivative 
\be\label{43A4}
\frac{\delta\Gamma}{\delta h_k(x)}=j^*_k(x).
\ee

We next want to show that the quantum effective action $\Gamma[h(x)]$ is invariant under fuzzy diffeomorphisms,
\be\label{43A5}
\delta_f\Gamma[h(x)]=\Gamma[h(x-\xi_f)]-\Gamma[h(x)]=0.
\ee
For smooth differentiable fields $h(x)$ and smooth displacements $\xi^\mu(x)$ this implies diffeomorphism symmetry of the effective action under the transformation \eqref{1}. For this purpose we write
\ba\label{22S}
\exp \big\{&-&\Gamma\big[h(x),j(x)\big]\big\}=\int\D\h(\tilde z)\exp\{-S\big[\h(x)\big]\nn\\
&+&\int_x\big(\h_k(x)-h_k(x)\big)j^*_k(x)+c.c\big\},
\ea
with $S\big[\h(x)\big]$ the diffeomorphism symmetric continuum action \eqref{x1}. The integrand on the r.h.s. is diffeomorphism symmetric if $h(x)$ and $\h(x)$ transform as scalars and $j(x)$ as a scalar density. The functional measure $\int\D\h(\tilde z)$ is invariant under diffeomorphisms, such that $\Gamma\big[h(x),j(x)\big]$ is diffeomorphism symmetric, and this extends to $\Gamma\big[h(x)\big]$. 

More in detail, we consider fuzzy diffeomorphisms and define the transformation of the source functions $\delta_fj(x)$ such that the r.h.s. of eq. \eqref{22Q} is invariant, 
\be\label{44A1}
\delta_fj(x)=-\partial_\mu\xi^\mu_f(x)j(x)-\xi^\mu_f(x)\partial_\mu j(x).
\ee
In the continuum limit $j(x)$ transforms as a scalar density with respect to the usual general coordinate transformations. With
\be\label{44A2}
\delta_fh(x)=-\xi^\mu_f(x)\partial_\mu h(x)
\ee
we compute the variation of the effective action 
\ba\label{44A3}
\delta_f\Gamma[h(x),j(x)]&=&\Gamma[h(x)+\delta_fh(x),j(x)+\delta_f j(x)]\nn\\
&&-\Gamma[h(x),j(x)].
\ea

In eq. \eqref{22S} the term $\sim\int_x hj^*$ is invariant under this fuzzy coordinate transformation, while the term $\int_x\h j^*$ also remains invariant if we shift simultaneously $\h(x)\to\h(x)+\delta_f \h(x)$. The shift in $\h$ affects the term involving $S\big[\h(x)\big]$. However, for any given lattice field $\h(\tilde z)$ we can achieve the shift $\h(x)\to\h(x-\xi_f)$ by a repositioning of the lattice points, {\em without} changing the lattice field $\h(\tilde z)$. (This holds precisely for $\xi^\mu_f$, being realized as interpolations of repositionings.) Since $S$ is diffeomorphism invariant, $\delta_fS[\h(x)]=0$, we conclude that the integrand, and therefore the integral, remains invariant under the transformation of the fields $h\to h+\delta_f h$, $j\to j+\delta_f j$. This means that $\Gamma$ is invariant if {\em only} the fields $h$ and $j$ are transformed. It is therefore a diffeomorphism invariant functional $\delta_f\Gamma[h(x),j(x)]=0$. 

The same argument applies to $W\big[j(x)\big]$,
\ba\label{22T}
W\big[j(x)\big]&=&\ln\int \D\h(\tilde z)\exp\big\{-S\big[\h(x)\big]\nn\\
&&+\int_x\big(\h_k(x)j^*_k(x)+c.c\big)\big\},
\ea
which is diffeomorphism symmetric when only $j(x)$ is transformed as a scalar density. This transformation property is compatible within
\be\label{22U}
\frac{\delta W[j]}{\delta j^*_k(x)}=h_k(x),
\ee
such that the solution of eq. \eqref{22U}, $j(x)\big[h(x)\big]$, transforms indeed as a scalar density if $h(x)$ is transformed as a scalar. Thus $\Gamma\big[h(x)\big]$ is invariant when one transforms only the field $h(x)$ to $h(x)+\delta_f h(x)$. 

The last step involves the continuum limit of smooth functions. Since an arbitrary general coordinate transformation $\xi^\mu(x)$ is well approximated by a suitable fuzzy coordinate transformation $\xi^\mu_f(x)$ we conclude that the quantum effective action is diffeomorphism symmetric in the continuum limit. 

In turn, the lattice version of the effective action 
\be\label{22V}
\Gamma[h(\tilde z)]=\sum_{\tilde y}V\y\bar L_\Gamma\big[h\y,\hp\mu h\y\dots\big]
\ee
is lattice diffeomorphism invariant. If not, $\bar L_\Gamma$ would involve $x$-dependent couplings. The $x$-dependence of the couplings would remain if we express $\bar L_\Gamma$ in terms of interpolating fields $h(x)$. This is in contradiction with the diffeomorphism symmetry which only transforms $h(x)$, and the general invariance with respect to repositioning if {\em both} couplings and lattice derivatives are transformed. The two invariances are compatible only for $x$-independent couplings. 

\section{Metric} 
\label{Metric}

So far we have used the coordinates $x^\mu$ only for the parametrization of a region of a continuous two-dimensional manifold. We have not used the notion of a metric and the associated ``physical distance''. (The physical distance differs from the coordinate distance $|x-y|$, except for the metric $g\mn=\delta\mn$.) The notion of a metric and the associated physical distance, topology and geometry can be inferred from the behavior of suitable correlation functions \cite{CWgeo}. Roughly speaking, for a euclidean setting the distance between two points $x$ and $y$ gets larger if a suitable properly normalized connected two-point function $G(x,y)$ gets smaller. This is how one world ``measure'' distances intuitively.

For the non-linear $\sigma$-model we may consider the two point function 
\be\label{M1}
G(x,y)=\kl \h_k(x)\h_k(y)\kr_c,
\ee
with $\h_k(x)$ suitable interpolating fields. Following ref. \cite{CWgeo} the metric is related to the behavior of $G(x,y)$ for $x\to y$ and can be defined as
\ba\label{M2}
g\mn(x)&=&\frac12\kl G\mn(x)+G^*\mn(x)\kr,\nn\\
G\mn(x)&=&\mu^{-2}_0\sum_k\hat\partial_\mu\h_k(x)\hat\partial_\nu\h_k(x).
\ea
The real normalization constant $\mu^{-1}_0$ has dimension of length such that $G\mn$ and $g\mn$ are dimensionless. For real $\h_k$ one has for the diagonal elements $g_{\mu\mu}\geq 0$. We generalize here the setting of ref. \cite{CWgeo} and also admit complex $\h_k$ where $g_{\mu\mu}$ can be negative. The signature of the metric is not defined a priori. Points where $\det\big(g\mn(x)\big)=0$ indicate singularities - either true singularities or coordinate singularities. More generally, the geometry and topology (e.g. singularities, identification of points etc.) of the space can be constructed from the metric \cite{CWgeo}. We also adapt to the lattice formulation and use in eq. \eqref{M2} interpolation derivatives. 

The metric is the central object in general relativity and appears in our setting as the expectation value of a suitable collective field. In practice, we do not need its relation to the behavior of correlation functions and we can simply take eq. \eqref{M2} as the {\em definition} of a metric. 	

For $x$ coinciding with the position of one of the cells $y_n=x_p(\tilde y_n)$ the interpolation derivative $\hat\partial_\mu\h_k(x)$ is given by the lattice derivative $\hat\partial_\mu\h_k(\tilde y)$. For these values of $x$ the field $G\mn(x)=G_{\mu\nu}\y$ can be expressed by lattice quantities
\ba\label{GA}
G\mn\y&=&\mu^{-2}_0\sum_k\hat\partial_\mu\h_k(\tilde y)\hat\partial_\nu\h_k(\tilde y)\nn\\
&=&\mu^{-2}_0a^{\tilde\mu}_\mu(x)a^{\tilde\nu}_\nu(x)G^{(L)}_{\tilde\mu\tilde\nu},
\ea
with ``lattice metric''
\be\label{M3}
G^{(L)}_{\tilde\mu\tilde\nu}=\frac13 p_{k,\tilde\mu}p_{k,\tilde\nu}
\ee
and 
\be\label{M4}
p_{k,0}=\h_k(\x4)-\h_k(\x1)~,~p_{k,1}=\h_k(\x3)-\h_k(\x2).
\ee
Here the functions $a^{\tilde \mu}_\mu(x)$ are defined implicitly by
\be\label{52A}
\hp\mu\h_k\y=a^{\tilde\mu}_\mu(x) p_{k,\mu}\y.
\ee
They depend on the positions $x_j$ of the points of a cell and are described in more detail in the appendix.

On the other hand, for $x\neq y_n$ the field $G\mn$ and therefore the metric $g\mn$ will depend on the specific interpolation prescription. This is again an expression of the fuzziness of space due to the lack of information beyond the lattice fields $\h_k(\tilde z)$. 

Similar to the lattice derivatives, the $x$-dependence of the metric arises only through the functions $a^{\tilde\mu}_\mu(x)$ which reflect the positioning of the lattice points. For interpolating functions $\h_k(x)$ transforming as scalars under fuzzy general coordinate transformations the metric \eqref{M2} transforms as a covariant second rank symmetric tensor with respect to fuzzy diffeomorphisms. In the continuum limit the interpolation derivatives in eq. \eqref{M2} are replaced by partial derivatives and the metric has the standard transformation property under general coordinate transformations. 

As a particular positioning we can use the regular lattice $x^\mu(\tilde z)=\Delta\tilde z^\mu$. This corresponds to a fixed choice of coordinates in general relativity. With this choice one has $V\y=2\Delta^2$ and
\be\label{M5}
a^{\tilde\mu}_\mu(x)=\frac{1}{2\Delta}\delta^{\tilde\mu}_\mu.
\ee 
Choosing $\mu^{-2}_0=4\Delta^2/3$, the collective field $G\mn$ in eq. \eqref{GA} coincides with the lattice metric $G^{(L)}\mn$ in eq. \eqref{M3}.

As an illustration, we may consider a particular configuration of lattice fields where for a given cell $\tilde y$ one has
\ba\label{M6}
\h_1(\x1)=\h_1(\x2)&=&\h_1(\x3)=\h_1(\x4)=H_1\y,\nn\\
\h_2(\x4)-\h_2(\x1)&=&H_2\y~,~\h_2(\x3)=\h_2(\x2),\nn\\
\h_3(\x3)-\h_3(\x2)&=&H_3\y~,~\h_3(\x4)=\h_3(\x1).
\ea
For this configuration the lattice metric $G^L\mn\y$ is diagonal,
\be\label{M7}
G_{00}=\frac13 H^2_2~,~G_{11}=\frac13 H^2_3~,~G_{01}=G_{10}=0,
\ee
while the cell action \eqref{6} is given by
\be\label{M8}
{\cal L}\y=\frac{\alpha}{12}H_1H_2H_3+c.c.
\ee
For appropriate choices of $H_1,H_2,H_3$ the combination $\alpha H_1H_2H_3$ can be real and negative, thus giving a substantial contribution to the functional integral. For example, this can be achieved for $\alpha>0$, real positive $H_3$, purely imaginary $H_1$ and $H_2$ with $H_1H_2<0$. In this case one finds a Minkowski signature of the lattice metric $G_{00}<0,G_{11}>0$. Euclidean signature obtains, for example, for real positive $H_2$ and real negative $H_1$. 

For the non-linear $\sigma$-model an $SO(3)$-rotation of the components $\h_k$ yields the same $G^L\mn$ and ${\cal L}\y$. There are, of course, many other configurations. For example, a cell with $\h_k(\x4)=\h_k(\x1)=-\h_k(\x3)=-\h_k(\x2)$ for one value of $k$ does not contribute to $G^L\mn\y$ or ${\cal L}\y$. If configurations of the type \eqref{M6} dominate one may expect an expectation value for the metric $g\mn$ of the type \eqref{M7}. 

Properties of the metric can often be extracted from symmetries. If the expectation values preserve lattice translation symmetry the metric $g\mn(x)$ will be independent of $x$. Invariance under a parity reflection implies $g_{01}=g_{10}=0$. Symmetry of the expectation values under lattice rotations would imply a flat euclidean metric $g_{00}=g_{11}$. A Minkowski metric $g\mn=\eta\mn$ requires that the expectation values violate the euclidean rotation symmetry.

\section{Effective action for gravity and gravitational field equations}
\label{Effective action for gravity and gravitational field equations}

The quantum effective action for the metric, $\Gamma[g_{\mu\nu}]$, can be constructed in the usual way by introducing sources for the collective field,
\ba\label{MB}
W[\tilde T]&=&\ln \int \D\h\exp\big\{-S+\int_xG^R\mn(x)\tilde T\hmn(x)\big\},\nn\\
G^R\mn&=&\frac12(G\mn+G^*\mn)\quad,\quad\frac{\delta W[\tilde T]}{\delta\tilde T\hmn(x)}=g\mn(x).
\ea
Solving formally for $\tilde T\hmn$ as a functional of $g\mn$, the effective action obtains by a Legendre transform
\be\label{MC}
\Gamma[g\mn]=-W+\int_x g\mn(x)\tilde T\hmn(x). 
\ee
The metric obeys the exact quantum field equation
\be\label{MD}
\frac{\delta\Gamma}{\delta g\mn(x)}=\tilde T\mn(x),
\ee
and we realize that $\tilde T\hmn$ can be associated to the energy momentum tensor $T\hmn$ by $\tilde T\hmn=\frac12\sqrt{g}T\hmn~,~g=|\det g\mn|$. 

The effective action $\Gamma[g\mn(x)]$ is invariant under fuzzy diffeomorphisms. This can be shown in analogy to the effective action for scalar fields in sect. \ref{Effective action}. We only sketch the proof here for the continuum limit of smooth fields.

Under a general coordinate transformation $\h_k(x)$ transforms as a scalar
\be\label{62}
\delta_\xi \h_k(x)=-\xi^\nu\partial_\nu \h_k(x).
\ee
This implies that $\partial_\mu \h_k$ and $G^R\mn$ transform as covariant vectors and second rank symmetric tensors, respectively. In consequence, $\tilde T\hmn$ transforms as a contravariant tensor density, with $T\hmn$ a symmetric second rank tensor. Thus $\int_xG^R\mn\tilde T\hmn$ and $\int_x g\mn\tilde T\hmn$ are diffeomorphism invariant, and $\Gamma[g\mn]$ is diffeomorphism invariant if $W[\tilde T]$ is diffeomorphism invariant. This is indeed the case for $\tilde T\mn$ transforming as a tensor, since in eq. \eqref{MB} the continuum action $S[\h(x)]$ is diffeomorphism invariant. For arbitrary interpolating fields (not necessarily smooth) the invariance of $\Gamma[g\mn(x)]$ under fuzzy diffeomorphisms follows if $\tilde T\hmn$ is transformed such that $\int_xG^R\mn\tilde T\hmn$ remains invariant.

For a given positioning of the lattice points and a given prescription for the interpolation procedure the functional integral \eqref{MB} is well defined and regularized for a finite number of lattice points. (This holds for arbitrary functions $\tilde T\hmn(x)$ provided the exponent in eq. \eqref{MB} remains bounded.) Therefore also $\Gamma\big[g\mn(x)\big]$ is a well defined functional that is, in principle, unambiguously calculable. (More precisely, this holds for all metrics for which the second equation \eqref{MB} is invertible.) A key question concerns the general form of the effective action $\Gamma[g\mn]$. If $\Gamma$ is diffeomorphism invariant and sufficiently local in the sense that an expansion in derivatives of $g\mn$ yields a good approximation for slowly varying metrics, then only a limited number of invariants as a cosmological constant or Einstein's curvature scalar $R$ contribute at long distances. The signature of the metric is not fixed a priori. For $g\neq 0$ the inverse metric $g\hmn$ is well defined. The existence of $g\hmn$ opens the possibility that $\Gamma[g\mn]$ also involves the inverse metric. 

This concludes our discussion of lattice diffeomorphism invariance for two-dimensional non-linear $\sigma$-models. Two dimensions are special for gravity, since the graviton does not propagate. However, all our constructions generalize to four dimensions as we will discuss in sect. \ref{Lattice diffeomorphism invariance in four dimensions}. Before doing so we will investigate lattice spinor gravity in two dimensions as a second type of model with a lattice diffeomorphism invariant action.

\section{Lattice spinor gravity}
\label{Lattice spinor gravity}

We next formulate quantum gravity in two dimensions based on fundamental fermions instead of bosons. The problem of boundedness of the action is completely absent for fermionic theories, where the partition function becomes a Grassmann functional integral. We may use for every lattice point two species, $a=1,2$, of two-component complex Grassmann variables $\varphi^a_\alpha(\tilde z),\alpha=1,2$, or equivalently eight real Grassmann variables $\psi^a_\gamma(\tilde z),\gamma=1\dots 4$, with $\varphi^a_1(\tilde z)=\psi^a_1(\tilde z)+i\psi^a_3(\tilde z)~,~\varphi^a_2(\tilde z)=\psi^a_2(\tilde z)+i\psi^a_4(\tilde z)$. The functional measure \eqref{3} is replaced by
\be\label{x2}
\int \D\psi=\prod_{\tilde z}\prod_\gamma\big (d\psi^1_\gamma(\tilde z)d\psi^2_\gamma(\tilde z)\big).
\ee
We introduce the bilinears (with Pauli matrices $\tau_k)$
\be\label{x3}
\h_k(\tilde z)=\varphi^a_\alpha(\tilde z)(\tau_2)_{\alpha\beta}(\tau_2\tau_k)^{ab}\varphi^b_\beta(\tilde z),
\ee
such that the action \eqref{6} contains now terms with six Grassmann variables. We keep the definitions \eqref{9}, \eqref{9A} and conclude that this fermionic action is lattice diffeomorphism invariant, leading to eq. \eqref{x1} in terms of Grassmann fields $\varphi^a_\alpha(x)$. In terms of $\psi^a_\gamma$ the action is an element of a real Grassmann algebra. 

The lattice derivatives for the Grassmann variables are defined similar to eq. \eqref{10} by the two relations
\ba\label{x4}
\varphi^a_\alpha(\x{j_1})-\tilde\varphi^a_\alpha(\x{j_2})=(x^\mu_{j_1}-x^\mu_{j_2})\hat\partial_\mu\varphi^a_\alpha\y
\ea
for $(j_1,j_2)=(4,1)$ and $(3,2)$. With 
\ba\label{x5}
&&\h_k(\x{j_1})-\h_k(\x{j_2})=
\big(\varphi^a_\alpha(\x{j_1})
+\varphi^a_\alpha(\x{j_2})\big)
(\tau_2)_{\alpha\beta}\nn\\
&&\qquad \qquad\times(\tau_2\tau_k)^{ab}\big(\varphi^b_\beta(\x{j_1})-\varphi^b_\beta(\x{j_2})\big),
\ea
and using reordering of the Grassmann variables, one obtains from eq. \eqref{6}
\ba\label{x6}
{\cal L}(y)&=&-8 i\alpha A\y\big(\varphi^a_\alpha(\x4)-\varphi^a_\alpha(\x1)\big)
(\tau_2)_{\alpha\beta}(\tau_2)^{ab}\nn\\
&&\times\big (\varphi^b_\beta(\x3)-\varphi^b_\beta(\x2)\big)+\dots,
\ea
with 
\be\label{x7}
A\y=\bar\varphi^1_1\y\bar\varphi^1_2\y\bar\varphi^2_1\y\bar\varphi^2_2\y,
\ee
and $\bar\varphi^a_\alpha\y$ the cell average. The dots indicate terms that do not contribute in the continuum limit. In terms of lattice derivatives \eqref{x4} one finds the action $S=\int d^2x\bar{\cal L}\y$,
\be\label{x8}
\bar\cl\y=-8i\alpha A\y\epsilon^{\mu\nu}\hat\partial_\mu\varphi^a_\alpha\y(\tau_2)_{\alpha\beta}(\tau_2)^{ab}\hat\partial_\nu
\varphi^b_\beta\y+\dots
\ee
For fixed spinor lattice derivatives \eqref{x4} the leading term \eqref{x8} is again lattice diffeomorphism invariant.

The continuum limit \eqref{x1} can be expressed in terms of spinors using $\partial_\mu\h_k(x)=2\varphi(x)\tau_2\otimes\tau_2\tau_k\partial_\mu\varphi(x)$, where the first $2\times 2$ matrix $E$ in $E\otimes F$ acts on spinor indices $\alpha$, the second $F$ on flavor indices $a$.   With
\be\label{N1}
F_{\mu\nu}=-A\partial_\mu\varphi\tau_2\otimes\tau_2\partial_\nu\varphi
\ee
one obtains
\be\label{N2}
S=4i\alpha\int d^2x\epsilon^{\mu\nu}F_{\mu\nu}+c.c.,
\ee
in accordance with eq. \eqref{x8}. Two comments are in order: (i) For obtaining a diffeomorphism invariant continuum action it is sufficient that the lattice action is lattice diffeomorphism invariant up to terms that vanish in the continuum limit. (ii) The definition of lattice diffeomorphism invariance is not unique, differing, for example, if we take fixed lattice derivatives \eqref{9A} for spinor bilinears or the ones \eqref{x4} for spinors. It is sufficient that the action is lattice diffeomorphism invariant for {\em one} of the possible definitions of lattice derivatives kept fixed. 

We finally note that $A$ and $F_{\mu\nu}$ are invariant under $SO(4,{\mathbbm C})$ transformations. This symmetry group rotates among the four complex spinors $\varphi^a_\alpha$, with complex infinitesimal rotation coefficients. For real coefficients, one has $SO(4)$, whereas other signatures as $SO(1,3)$ are realized if some coefficients are imaginary. The continuum action \eqref{N2} or \eqref{x1} exhibits a local $SO(4,{\mathbbm C})$ gauge symmetry. A subgroup of $SO(4,{\mathbbm C})$ is the two-dimensional Lorentz group $SO(1,1)$. The action \eqref{6} is therefore a realization of lattice spinor gravity \cite{A} in two dimensions.

\section{Lattice diffeomorphism invariance in four dimensions} 
\label{Lattice diffeomorphism invariance in four dimensions}

We generalize the lattice to four dimensions with integers $(\tilde z^0,\tilde z^1,\tilde z^2,\tilde z^3)$ and $\sum_\mu\tilde z^\mu$ odd. Each cell located at $\tilde y=(\tilde y^0,\tilde y^1,\tilde y^2,\tilde y^3)$, with $\sum_\mu\tilde y^\mu$ even, contains eight lattice points, located at the nearest neighbors of $\tilde y$ at $\tilde y\pm v_0,\dots\tilde y\pm v_3$, with $(v_\mu)^\nu=\delta^\nu_\mu$. We double the number of degrees of freedom with bosons $\h^+_k$ and $\h^-_k$, or fermions $\varphi^a_{+\alpha},\varphi^a_{-\alpha}$, using 
\be\label{N3}
\h^\pm_k=\varphi_\pm C_\pm\otimes\tau_2\tau_k\varphi_\pm~,~C_\pm=\pm\tau_2.
\ee
We further introduce
\ba\label{N4}
{\cal F}^\pm_{\mu\nu}\y&=&\frac{1}{12}\epsilon^{klm}\bar\h^\pm_k\y\nn\\
&\times&\big[\h^\pm_l(\tilde y+v_\mu)-\h^\pm_l(\tilde y-v_\mu)\big]\nn\\
&\times&\big[\h^\pm_m(\tilde y+v_\nu)-\h^\pm_m(\tilde y-v_\nu)\big],
\ea
with $\bar\h\y$ the cell average. (Note that $\cl(y)$ in eq. \eqref{6} obeys ${\cal L}=\alpha{\cal F}_{01}+c.c.$). A lattice diffeomorphism invariant action in four dimensions can be written as
\be\label{N5}
S=\frac{\tilde\alpha}{24}\sum_{\tilde y}\epsilon^{\mu\nu\rho\sigma}{\cal F}^+_{\mu\nu}{\cal F}^-_{\rho\sigma}+c.c.
\ee

We define the lattice derivatives by the four relations
\be\label{N6}
\h(\tilde y+v_\nu)-\h(\tilde y-v_\nu)=
(x^+_\nu-x^-_\nu)^\mu\hat\partial_\mu\h\y,
\ee
where $x^\pm_\nu=x_{p}(\tilde y\pm v_\nu)$. With $\Delta_\nu=(x^+_\nu- x^-_\nu)/2$ the cell volume amounts to 
\ba\label{N7}
V\y&=&2\epsilon_{\mu\nu\rho\sigma}\Delta^\mu_0\Delta^\nu_1\Delta^\rho_2\Delta^\sigma_3\nn\\
&=&\frac{1}{12}
\epsilon_{\mu\nu\rho\sigma}\epsilon^{\mu'\nu'\rho'\sigma'}\Delta^\mu_{\mu'}\Delta^\nu_{\nu'}\Delta^\rho_{\rho'}
\Delta^\sigma_{\sigma'}.
\ea
Using $\int d^4x=\sum_{\tilde y}V\y$ one finds indeed that the action does not depend on the positioning of the lattice points,
\be\label{N8}
S=\frac{\tilde\alpha}{3}\int d^4x\epsilon^{\mu\nu\rho\sigma}\hat{\cal F}^+_{\mu\nu}\hat{\cal F}^-_{\rho\sigma}+c.c.,
\ee
with 
\be\label{N9}
\hat{\cal F}^\pm_{\mu\nu}\y=\frac{1}{12}\epsilon^{klm}\bar\h^\pm_k\y\hat\partial_\mu\h^\pm_l\y\hat\partial_\nu\h^\pm_m\y.
\ee
The continuum limit $\bar \h\to\h,\hat\partial_\mu\to\partial_\mu$, is diffeomorphism invariant due to the contraction of the partial derivatives with the $\epsilon$-tensor. 

For bosons one may employ real or complex fields $\h^\pm_k$ and define a non-linear $\sigma$-model by the condition
\be\label{32A}
(\h^\pm_k)^*\h^\pm_k=1.
\ee
For complex fields the non-linear $\sigma$-model is invariant under a global $SU(3)_+\times SU(3)_-$ symmetry acting on the index $k$ of $\h^+_k$ and $\h^-_k$, respectively. For complex bosons it is possible to identify $(\h^+_k)^*$ with $\h^-_k$ or $(\h^-_k)^*$, thus reducing the ``flavor symmetry'' to $SU(3)$. For real bosons $\h_k$ the symmetry is reduced to $SO(3)\times SO(3)$.

For spinor gravity one uses eq. \eqref{N3} and defines
\be\label{N10}
F^\pm_{\mu\nu}=A^\pm D^\pm_{\mu\nu},
\ee
with $A^\pm$ defined as in eq. \eqref{x7} and 
\be\label{N11}
D^\pm_{\mu\nu}=-\partial_\mu\varphi_\pm\tau_2\otimes\tau_2\partial_\nu\varphi_\pm.
\ee
This yields for the continuum limit
\be\label{N12}
\hat{\cal F}^\pm_{\mu\nu}\to\pm 4iF^\pm_{\mu\nu}.
\ee
The lattice action for spinor gravity proposed in ref. \cite{A} corresponds to eq. \eqref{N5} if we replace in eq. \eqref{N4} the cell average $\bar\h\y$ by the plane average $\big[\h(\tilde y+v_\mu)+\h(\tilde y-v_\mu)+\h(\tilde y+v_\nu)+\h(\tilde y-v_\nu)\big]/4$. The difference vanishes in the continuum limit. 

For spinor gravity $F^+_{\mu\nu}$ is invariant under $SO(4,{\mathbbm C})_+$ transformations acting on $\varphi_+$, while $F^-_{\mu\nu}$ is invariant under $SO(4,{\mathbbm C})_-$ transformations acting on $\varphi_-$. Since $A_+A_-$ involves already the maximal number of eight different complex spinors $\varphi^a_{+\alpha},\varphi^a_{-\alpha}$ at a given point, all inhomogeneous terms vanish for local transformations \cite{A}. Thus the action \eqref{N8} is invariant under local $SU(2,{\mathbbm C})_+\times SU(2,{\mathbbm C})_-\times SU(2,{\mathbbm C})_L\times SU(2,{\mathbbm C})_R$ transformations, where $\varphi_+$ transform as $(2,1,2,1)$ and $\varphi_-$ as $(1,2,1,2)$. Here $SO(4,{\mathbbm C})_+=SU(2,{\mathbbm C})_+\times SU(2,{\mathbbm C})_L$ and $SO(4,{\mathbbm C})_-=SU(2,{\mathbbm C})_-\times SU(2,{\mathbbm C})_R$, and $SU(2,{\mathbbm C})$ denotes the $SU(2)$-transformations with three arbitrary complex infinitesimal parameters, equivalent to six real parameters. We can identify the generalized Lorentz group with $SO(4,{\mathbbm C})=SU(2,{\mathbbm C})_+\times SU(2,{\mathbbm C})_-$, such that $\varphi_+$ and $\varphi_-$ transform as two inequivalent Weyl spinors with opposite handedness. The standard Lorentz group is the $SO(1,3)$ subgroup of $SO(4,{\mathbbm C})$. Choosing real transformation parameters, the subgroup $SU(2)_L\times SU(2)_R$ can be identified with a local gauge symmetry.

The continuum limit of the action of lattice spinor gravity, as defined by eqs. \eqref{N8}, \eqref{N12}, namely
\ba\label{N13} 
S&=&\frac{16\tilde\alpha}{3}\int d^4xA^+A^-D+c.c.,\nn\\
D&=&\epsilon^{\mu\nu\sigma\rho}D^+_{\mu\nu}D^-_{\rho\sigma},
\ea
differs from earlier versions of spinor gravity \cite{HCW}, \cite{CWSG}, and also from first attempts to formulate a diffeomorphism invariant action in terms of spinors without employing a metric \cite{Aka,Ama,Den}. It realizes local Lorentz symmetry and has the property that the difference of signature between time and space only arises from the dynamics rather than being imposed in the action. In this respect it resembles the higher dimensional model of ref. \cite{CWLL}. 

The action \eqref{N13} can be expressed in terms of suitable vierbein bilinears
\be\label{N14}
\tilde E^m_\mu=\varphi C_+\gamma^m_M\otimes V\partial_\mu\varphi \pm \varphi C_-\gamma^m_M\otimes V\partial_\mu\varphi, 
\ee
with $\varphi=(\varphi_+,\varphi_-)$, $\gamma^m_M$ the usual $4\times 4$ Dirac matrices and $V$ suitable $2\times 2$ flavor matrices. (This reformulation will be described in a separate publication, see ref. \cite{CWMS} for conventions.)  The vierbein bilinears are vectors with respect to general coordinate transformations and vectors with respect to global Lorentz transformations. Diffeomorphism symmetry and local Lorentz symmetry of the action guarantee that only invariants similar to Cartan's formulation of general relativity in terms of vierbeins \cite{Car} can appear in this expression of the continuum action. The presence of several vierbein-type objects (e.g. for different $V$) induces, however, new interesting features of a mixing between geometrical and flavor aspects. For a suitable choice of the fermion bilinear $\tilde E^m_\mu$ one may associate its expectation value with the vierbein of general relativity. 

From the vierbein bilinear we can construct a collective field for the metric as
\be\label{78A}
G\mn=\mu^{-2}_0\tilde E^m_\mu\tilde E^n_\nu\eta_{mn}.
\ee
(The special role of the symmetry $SO(1,3)$ suggested by the presence of $\eta_{mn}=diag(-1,1,1,1)$ is only apparent. One could define a euclidean vierbein by multiplying $\tilde E^0_\mu$ by a factor $i$, thus replacing $\gamma^m_M$ by euclidean Dirac matrices. In terms of those eq. \eqref{78A} involves $\delta_{mn}$ instead of $\eta_{mn}$.) We may again define the metric by eq. \eqref{M2} and define the diffeomorphism invariant effective action $\Gamma[g\mn]$ similar to eqs. \eqref{MB}, \eqref{MC}. We note in this respect that there are several collective fields transforming as symmetric second rank tensors which are candidates for the metric. We first have different possible choices $V$. We could also use $g\mn=\mu^{-2}_0\kl \tilde E^m_\mu\kr\kl\tilde E^n_\nu\kr\eta_{mn}$ instead of eq. \eqref{78A}. Finally we could employ a lattice metric of the type \eqref{M3}, with a suitable choice of $\sigma_1,\sigma_2=(+,-)$,
\be\label{A10-2}
G^{(L)}_{\tilde\mu\tilde\nu}=\frac16 \big(p^{\sigma_1}_{k,\tilde\mu}p^{\sigma_2}_{k,\tilde\nu}+(\mu\leftrightarrow\nu)\big),
\ee
with 
\be\label{A10-2a}
p^\sigma_{k,\tilde\mu}=\h^\sigma_k(\tilde y+v_{\tilde\mu})-\h^\sigma_k(\tilde y-v_{\tilde\mu}).
\ee
The metric obtains then by a generalization of eq. \eqref{GA}. This metric can be employed for bosons as well. (For fermions the different candidates are not all independent, due to the possibility to reorder the fermions.) We could even discuss effective models with several distinct metrics. We expect, however, that in general only one particular linear combination will remain massless, whereas the others are massive. The massless mode can be associated with the physical metric and the graviton. 

For a vanishing energy momentum tensor one expects that the expectation values of fields are invariant under some of the symmetries of the action, but not under all of them. (Certain symmetries will be spontaneously broken.) The conserved symmetries determine the form of the metric to a large extent. This may be illustrated by a discussion of the consequences of discrete symmetries in case of spinor gravity or for complex bosons $\h_k$. Let us distinguish a ``time coordinate'' $x^0$ from the three ``space coordinates'' $x^k$. A reflection of one of the space coordinates, $x^k\to-x^k$, has to be accompanied by a transformation $\h\to{\cal P}\h$ acting on the ``internal indices'' of $\h$. This is needed in order to leave the action invariant. The required condition $S[{\cal P}\h]=-S[\h]$ can be achieved by changing the sign of an odd number of components $\h_k$, for example $\h^-_k\to-\h^-_k,\h^+_k\to\h^+_k$. The transformation $\h\to{\cal P}\h$ leaves $g\mn$ invariant, such that invariance under space reflections, $\partial_k\to-\partial_k$, implies that the metric is diagonal. If we further assume that the expectation values are invariant under $\pi/2$-rotations among the space coordinates one infers $g_{kl}=a\delta_{kl}$. Lattice translation symmetry of expectation values implies that $a$ is independent of $x$. Furthermore, the action is also invariant under ``diagonal reflections'' $x^0\leftrightarrow x^k$ which are accompanied by an internal transformation $\h\to\D\h$, with $S[\D\h]=-S[\h]$. An example is $\D\h=i\h$, which changes the sign of the metric. Then invariance under the diagonal reflections, $\partial_0\leftrightarrow \partial_k$ implies, $g_{00}=-g_{kk}$. The combined symmetries therefore lead to flat Minkowski space, $g\mn=a\eta\mn$. It remains to be seen if the non-linear $\sigma$-model or spinor gravity with action \eqref{N5} lead indeed to a ``ground state'' of this type. 

\section{Conclusions} 
\label{Conclusions}

We have formulated the property of lattice diffeomorphism invariance for lattice models of quantum gravity that do not involve a fundamental metric or geometric degrees of freedom. The functional integrals are defined in terms of variables at discrete lattice points which behave as scalars under diffeomorphisms. We have constructed lattice diffeomorphism invariant actions for discrete bosons, continuous bosons and fermions, both for two and four dimensions. We have shown that lattice diffeomorphism invariance implies diffeomorphism symmetry of the continuum action and the quantum effective action. 

The metric $g\mn$ arises as the expectation value of a suitable collective field. The effective action for the metric is diffeomorphism invariant, which implies that the gravitational field equations are covariant with respect to general coordinate transformations. For a finite number of lattice points the functional integral is fully regularized. The effective action $\Gamma[g\mn]$ \eqref{MC} for the metric is then well defined. The correlation functions of the metric are, in principle, computable - even though this may be hard in practice. This also holds for the quantum field equations which obtain from the variation of the effective action. Consistent geometries correspond to solutions of these field equations with an energy momentum tensor as a source. 

The bosonic Ising type or non-linear $\sigma$-models are suitable for numerical lattice simulations. Of key interest is a computation of expectation value and correlation functions of the metric. These can be compared to the ones that follow from a diffeomorphism invariant action in a derivative expansion. In particular, in four dimensions one expects the long range correlations appropriate for a massless graviton.  For real fields $\h_k$ the signature of the metric is necessarily euclidean. For the non-linear $\sigma$ model with complex bosons $\h_k$ we may investigate the case of Minkowski signature as well. A numerical evaluation of Grassmann integrals is more involved, in particular for the non-Gaussian action \eqref{N5}. Analytical approximations for the quantum effective action could be based on the $2PI$-formulation \cite{D13a} for the bosonic effective action \cite{D13b}, as sketched in ref. \cite{D13c}. This requires the solution of gap equations in the presence of bosonic background fields as the metric. 

A priori, all solutions of the gravitational field equation \eqref{MD} are viable if we include appropriate sources and boundary terms in the definition of the functional integral. This includes cosmological solutions. For an effective action $\Gamma[g\mn]$ that is invariant under general coordinate transformations and sufficiently local, one expects for long distances the validity of a derivative expansion with cosmological constant, Einstein's curvature scalar and higher order terms. If this turns out to be the case the lattice diffeomorphism invariant actions realize the construction of regularized quantum gravity.

\section*{APPENDIX: RESTRICTED LATTICE DIFFEOMORPHISMS}
\renewcommand{\theequation}{A.\arabic{equation}}
\setcounter{equation}{0}

In the main text we have discussed diffeomorphism transformations on the level of interpolating functions. Differentiable interpolating fields obey the usual rules for transformations of scalars, densities, covariant vectors and tensors and so on. The lattice regularization imposes only the restriction that we consider only those infinitesimal displacements $\xi^\mu_f(x)$ that can be obtained as interpolating functions of a repositioning of the lattice points on the manifold. In the continuum limit of smooth fields this restriction plays no role. The lattice functions $\h(\tilde z)$ are not affected by diffeomorphism transformations. This guarantees the diffeomorphism invariance of a functional measure defined in terms of $\h(\tilde z)$. General coordinate transformations act only on the interpolating fields and correspond to a change of positioning of lattice points $x_n$ in ${\mathbbm R}^2$. 

In this appendix we investigate lattice diffeomorphism transformations that act directly on the discrete lattice variables, rather than on interpolating functions. For simplicity of the presentation we concentrate on two dimensions. Again, infinitesimal lattice diffeomorphism transformations correspond to an infinitesimal change $x_p(\tilde z)\to x_p(\tilde z)+\xi_p(\tilde z)$ of the positioning of the lattice points. We study here the change of the discrete lattice derivatives $\hat\partial_\mu\h(\tilde y)$ resulting from such a repositioning. For ``restricted lattice diffeomorphisms'' we keep the cell averages $\h(\tilde y)$ fixed. If we keep also the positions $x_p(\tilde y)$ of the cells fixed (for example on a regular lattice $x_p(\tilde y)=\Delta\tilde y)$ the geometric meaning of the restricted diffeomorphisms can be visualized as moving the the positions of the lattice points $\tilde z$ at fixed position for $\tilde y$. This changes the shape and volume of each cell, but not its position. (The cell position discussed in this appendix may differ from the one used for some given interpolating description.) 

The transformation of the lattice derivatives follows from the change of positions $x^\mu_j$ in eq. \eqref{9A}, while the cell average \eqref{9} is kept fixed. Similar to the usual general coordinate transformations one can define the notion of ``lattice vectors'' and ``lattice tensors'' with respect to these restricted transformations. For a scalar lattice function $\tilde f(\tilde z)$ the lattice derivatives transform as a lattice vector. Defining
\be\label{22A}
p_0\y=\tilde f(\x4)-\tilde f(\x1)~,~p_1\y=\tilde f(\x3)-\tilde f(\x2)
\ee
and
\be\label{22B}
d^\mu_0=x^\mu_4-x^\mu_1~,~d^\mu_1=x^\mu_3-x^\mu_2,
\ee
we can write the lattice derivative as
\be\label{22C}
\hat\partial_\mu f\y=a^{\tilde\mu}_\mu(x) p_{\tilde \mu},
\ee
with 
\be\label{22D}
a^{\tilde \mu}_\mu(x)=\frac{1}{2V\y}\epsilon\mn\epsilon^{\tilde\mu\tilde \nu}d^\nu_{\tilde\nu}.
\ee
The lattice derivative depends on the positions of the lattice points through the $x_j$-dependent coefficients $a^{\tilde\mu}_\mu(x)$ which multiply the differences of lattice variables $p_{\tilde \mu}$. We observe the relations
\ba\label{22DA}
&&V\y=\frac14\epsilon\mn\epsilon^{\tilde\mu\tilde\nu} d^\mu_{\tilde\mu}d^\nu_{\tilde\nu},\nn\\
&&\epsilon\mn d^\mu_{\tilde\mu}d^\nu_{\tilde\nu}=2V\epsilon_{\tilde\mu\tilde\nu},
\ea
and
\ba\label{22E}
&&a^{\tilde\mu}_\mu d^\mu_{\tilde\mu}=2,\nn\\
&&\epsilon\hmn a^{\tilde\mu}_\mu a^{\tilde\nu}_\nu=\frac{1}{4V^2}\epsilon\mn\epsilon^{\tilde\mu\tilde\rho}
\epsilon^{\tilde\nu\tilde\sigma}d^\mu_{\tilde\rho}d^\nu_{\tilde\sigma},\nn\\
&&\epsilon\hmn\epsilon_{\tilde\mu\tilde\nu}a^{\tilde\mu}_\mu a^{\tilde\nu}_\nu=\frac1V.
\ea

With respect to infinitesimal repositionings one has 
\be\label{22F}
\delta_p d^\mu_0=\xi^\mu_p(\x4)-\xi^\mu_p(\x1)~,~
\delta_p d^\mu_1=\xi^\mu_p(\x3)-\xi^\mu_p(\x2),
\ee
and infers for the change of the cell volume
\be\label{22G}
\delta_pV\y=\hat\partial_\mu\xi^\mu_p\y V\y.
\ee
Here $\hat\partial_\mu\xi^\nu_p\y$ is formed with the standard prescription \eqref{22C} for lattice derivatives, using for $p_{\tilde \mu}$ differences of $\xi^\nu_p(\tilde z)$. For the functions $a^{\tilde\mu}_\mu(x)$ a change of positioning results in
\be\label{22H}
\delta_p a^{\tilde\mu}_\mu(x)=-\hat\partial_\mu\xi^\nu_p\y a^{\tilde\mu}_\nu(x).
\ee
Since $p_{\tilde\mu}$ does not depend on the positioning one obtains from eq. \eqref{22C}
\be\label{22I}
\delta_p\hat\partial_\mu f\y=-\hat\partial_\mu\xi^\nu_p\y\hat\partial_\nu f\y.
\ee

The transformation property $\delta_p v_\mu\y=-\hat\partial_\mu\xi^\nu_p\y v_\nu\y$ defines a lattice vector with respect to restricted lattice diffeomorphism transformations. For smooth enough lattice functions $v_\mu\y$ and $\xi^\nu_p\y$ the lattice derivatives become standard partial derivatives of interpolating functions. If we extend the discussion beyond the restricted transformations the repositioning also associates to the cell $\tilde y$ a new position $x_p\y+\xi_p\y$. From $\delta_\xi v_\mu(x)=\delta_p v_\mu\y+v_\mu(x-\xi)-v_\mu(x)$ one recovers for the associated interpolating functions the standard transformation property of a covariant vector with respect to general coordinate transformations
\be\label{22J}
\delta_\xi v_\mu=-\partial_\mu\xi^\nu v_\nu-\xi^\nu\partial_\nu v_\mu.
\ee
We conclude that the transformation of lattice vectors under restricted lattice diffeomorphism transformations translates for the associated interpolating functions to the transformation law of covariant vectors under general coordinate transformations. This holds in the continuum limit of smooth $v_\mu$ and $\xi^\nu$.

Tensors with respect to restricted lattice diffeomorphisms (``lattice tensors'') can be obtained in a standard way from multiplication of lattice vectors. For example, $\hat\partial_\mu f\y\hat\partial_\nu g\y+\hat\partial_\nu f\y\hat\partial_\mu g\y$ transforms as a second rank symmetry lattice tensor. This is the transformation of the metric field $G_{\mu\nu}$ in eq. \eqref{GA}. The antisymmetric lattice tensor 
\ba\label{22K}
A\mn\y &=&\hat\partial_\mu f\y\hat\partial_\nu g\y-\hat\partial_\nu f\y\partial_{\hat\mu} g\y\nn\\
&=&\epsilon\mn\epsilon^{\rho\sigma}\hat\partial_\rho f\y\hat\partial_\sigma g\y
\ea
transforms in the same way as $1/V\y$,
\be\label{22L}
\delta_p\big(\epsilon^{\rho\sigma}\hat\partial_\rho f\y\hat\partial_\sigma g\y\big)=-
\hp\mu\xi^\mu_p\big(\epsilon^{\rho\sigma}\hp\rho f\y\hp\sigma g\y\big),
\ee
such that $V\y A\mn\y$ is invariant. 

As a consequence, the ``cell action'' $\hl\y$ in eq. \eqref{13} is invariant under restricted lattice diffeomorphisms. This is another facet of lattice diffeomorphism invariance. In fact, every lattice action can be expressed in terms of suitable cell averages and lattice derivatives. In general, however, the couplings multiplying different terms will depend on the coordinates $x^\mu_p(\tilde z)$ of the lattice points. Since the original lattice action does not ``know'' about the positioning, the expression of the action in terms of cell averages and derivatives, $\hl\y$, is invariant under a repositioning if {\em both} the cell averages and associated cell derivatives are transformed by lattice diffeomorphism transformations, {\em and} the couplings are transformed according to their dependence on $x_p$. For a lattice diffeomorphism invariant model $\hl\y$ transforms as $1/V\y$ if {\em only} the fields $\h\y$ and associated derivatives $\hat\partial_\mu\h\y$ are transformed. In this case $\hat L\y$ has to transform as $V\y$ if the couplings are transformed, while $\h\y$ and $\hat\partial_\mu\h\y$ are kept fixed. One can then write $\hat{\cal L}\y=V\y\bar{\cal L}\y$ and conclude that $\bar{\cal L}\y$ does not involve any couplings that depend on $x_p$. This is precisely the characterization of lattice diffeomorphism invariance in sect. \ref{Lattice diffeomorphism invariance in two dimensions}.

At this point a general principle for the construction of lattice diffeomorphism invariant actions becomes visible. Since $\delta_p\h\y=0$, all lattice actions with 
\ba\label{36A}
\bar\cl\y=f\big(\h\y\big)\epsilon\hmn\hp\mu\h_l\y\hp\nu\h_m\y
\ea
transform $\sim V\y^{-1}$ under lattice diffeomorphism transformations for arbitrary functions $f$. The associated continuum limit replaces $\h\y$ by the interpolating field $\h(x)$, and $\hp\mu\tilde\h\y\to\hat\partial_\mu\h(x)$. The expression of the associated $\hl\y=\bar\cl\y V\y$ in terms of the lattice field $\h(\tilde z)$ is straightforward and does not involve the positions of the lattice points. This construction is easily generalized to four dimensions or any other number of dimensions. In $d$ dimensions it formulates a diffeomorphism invariant action as an integral over a $d$-form. It is not known if other forms of lattice diffeomorphism invariant actions exist that do not follow this construction. 

We conclude that lattice diffeomorphism invariant actions are those for which $\bar{\cal L}\y$ transforms as
\be\label{37A}
\delta_p\bar\cl\y=-\hat\partial_\mu\xi^\mu_p\y\bar\cl\y
\ee
under restricted lattice diffeomorphism transformations. Here we recall that $\delta_p$ is taken for fixed $\tilde y$, such that $\delta_p\h\y=0$ while $\partial_p(\hat\partial_\mu\h)$ is given by eq. \eqref{22I}. 

Using the congruent action interpolation one obtains the continuum action 
\be\label{37B}
S=\int_x\bar\cl (x),
\ee
where the lattice derivative $\hat\partial_\mu f\y$ is replaced by the ``interpolation derivative'' $\hat\partial_\mu f(x)$ defined by eq. \eqref{A1} and obeying the relation \eqref{A3}, and $\h\y$ is replaced by $\h(x)$. This continuum action coincides with the lattice action. For the continuum action we have to take into account the volume factor in eq. \eqref{IA} such that with eq. \eqref{22G} the variation under a (restricted) repositioning of lattice points becomes 
\be\label{37C}
\delta_p S=\int_x\big(\partial_\mu\xi^\mu_p(x)\bar\cl(x)+\delta_p\bar\cl(x)\big).
\ee

For a lattice diffeomorphism invariant action obeying eq. \eqref{37A} the continuum action is invariant, $\delta_p S=0$. We emphasize that $\delta_p\bar\cl(x)$ is a well defined operation for any $\bar\cl(x)$ that can be written as a sum of terms for which each term is a polynomial in derivatives $\hat\partial_\mu f(x)$, multiplied by an arbitrary function of $f(x)$. (Generalizations to several functions $f_a(x)$ are obvious.) It simply transforms all derivatives of functions according to eq. \eqref{22I}, without any further change,
\be\label{37D}
\delta_p\big(\hat\partial_\mu f(x)\big)=-\hat\partial_\mu\xi^\nu_p(x)\partial_\nu f(x).
\ee

For congruent action interpolations the use of (continuous) functions $\h(x)$ that are defined for all points of a continuous manifold is a simple rewriting of the lattice model in a different picture. In this picture the action is given as a functional of $\h(x)$. In general, different positionings of the lattice points result in different pictures, characterized by different continuum actions. Restricted lattice diffeomorphism transformations correspond to a repositioning of lattice points on a manifold and therefore to a switch between different pictures. For a lattice diffeomorphism invariant action all pictures for arbitrary positionings are the same.

We next turn to the restricted lattice diffeomorphism transformations of sources. We start with sources for cell averages, $J\y=\frac14\Sigma_j J\big(\x j\y\big)$. If $J\y$ transforms as a scalar, the ratio $j\y=J\y/V\y$ transforms as a scalar density, with 
\be\label{22M}
\delta_pj\y=-\hp\nu\xi^\nu_p\y j\y,
\ee
or, for smooth lattice fields,
\be\label{22N}
\delta_\xi j(x)=-\partial_\nu\big(\xi^\nu(x)j(x)\big).
\ee
Lattice diffeomorphism invariance is realized if $\bar \cl$ transforms as a scalar density when only $\hp\mu\h\y$ are transformed according to restricted lattice diffeomorphisms. This extends to a source term for cell averages,
\be\label{39A1}
\bar\cl_j\y=-\big(\h^*_k\y j_k\y+\h_k\y j^*_k\y\big),
\ee
provided the sources transform according to eq. \eqref{22M}.

We further notice that we can consider $a^{\tilde\mu}_\mu(x)$ as a regular $2\times 2$ matrix, since the definition implies $\det a=1/2V>0$. The inverse matrix $b^\mu_{\tilde\mu}(x)$ obeys
\be\label{37E}
a^{\tilde\mu}_\mu b^\nu_{\tilde\mu}=\delta^\nu_\mu~,~b^\mu_{\tilde\mu}a^{\tilde\nu}_\mu=\delta^{\tilde\nu}_{\tilde\mu},
\ee 
and transforms as
\be\label{37F}
\delta_p b^\mu_{\tilde\mu}(x)=b^\nu_{\tilde\mu}(x)\hp\nu\xi^\mu_p\y.
\ee
This is the transformation property of a contravariant vector. Contraction with any invariant object $\tilde w^{\tilde\mu}\y$ yields a contravariant vector 
\be\label{37G}
w^\mu \y=b^\mu_{\tilde \mu}(x)\tilde w^{\tilde\mu}\y,
\ee
which transforms as
\be\label{37H}
\delta_p w^\mu\y=w^\nu\y\hat\partial_\nu\xi^\mu_p\y.
\ee
The product of a covariant and a contravariant vector is invariant
\be\label{37I}
\delta_p(v_\mu w^\mu)=0.
\ee

\ba\label{37J}
\quad \text{With } \tilde w^0&=&J(\x4)-J(\x1)~,~\tilde w^1=J(\x3)-J(\x2), \text{ and}\nn\\
v_\mu&=&\hat\partial_\mu\h\y=a^{\tilde\mu}_\mu p_{\tilde\mu},\nn\\
w^\mu&=&\hat\partial^\mu J\y=b^\mu_{\tilde\mu}\tilde w^{\tilde \mu},
\ea
one finds
\ba\label{37K}
\hat\partial_\mu\h\hat\partial^\mu J&=&
\tilde p_{\tilde\mu}\tilde w^{\tilde\mu}=
[\h(\x4)-\h(\x1)][J(\x4)-J(\x1)]\nn\\
&&+[\h(\x3)-\h(\x2)][J(\x3)-J(\x2)].
\ea
Contravariant lattice vectors $\partial^{\hat\mu}J\y$ related to sources introduce new possibilities for constructing lattice diffeomorphism invariants in $\Gamma[h,j]$ and therefore also for the effective action $\Gamma[h]$.

\end{document}